\documentclass[a4paper,11pt]{article}
\pdfoutput=1 

\usepackage{jcappub} 
\usepackage[T1]{fontenc} 

\usepackage{booktabs}

\title{Testing Lorentz Invariance of Gravity in the Standard Model Extension with GWTC-3}

\author[a, b]{Rui Niu}
\author[c, d]{Tao Zhu}
\author[a, b]{Wen Zhao}

\affiliation[a]{CAS Key Laboratory for Researches in Galaxies and Cosmology, Department of Astronomy, University of Science and Technology of China, Chinese Academy of Sciences, Hefei, Anhui 230026, China}
\affiliation[b]{School of Astronomy and Space Sciences, University of Science and Technology of China, Hefei 230026, China}
\affiliation[c]{Institute for Theoretical Physics and Cosmology, Zhejiang University of Technology, Hangzhou, 310032, China}
\affiliation[d]{United Center for Gravitational Wave Physics (UCGWP), Zhejiang University of Technology, Hangzhou, 310032, China}

\emailAdd{nrui@mail.ustc.edu.cn}
\emailAdd{zhut05@zjut.edu.cn}
\emailAdd{wzhao7@ustc.edu.cn}

\abstract{Successful detection of gravitational waves has presented a new avenue to explore the nature of gravity. With the cumulative catalog of detected events, we can perform tests on General Relativity from various aspects with increasing precision.
In this work, we focus on Lorentz symmetry during propagation of gravitational waves. 
Considering the dispersion relation in the gauge-invariant linearized gravity sector of the Standard-Model Extension, the anisotropy, birefringence, and dispersion effects will be induced during propagation of gravitational waves because of the Lorentz violating modification, and cause dephasings in waveform received by detectors.
With the distorted waveform, we perform full Bayesian inference with confident events in the last gravitational wave catalog. We consider two cases associated with the lowest mass dimension $d=5,6$ which are supposed to have the most significant effects, and place the constraints on the expansion coefficients characterizing the Lorentz violating behavior which have 16 independent components for $d=5$ and 18 components for $d=6$.
We do not find any evidence for Lorentz violation in the gravitational wave data, the constraints on the coefficients are on the order of $10^{-15}{\rm m}$ for $d=5$ and $10^{-10}{\rm m^2}$ for $d=6$ respectively.
}

\begin{document}
\maketitle
\flushbottom

\section{Introduction} \label{sec_intro}
General Relativity (GR) is the most successful theory for describing  gravity phenomena, and has passed various experimental tests on different scales with incredible precision \cite{Berti2015,Will2014,Hoyle2001,Adelberger2001,Jain2010,Koyama2016,Clifton2012,Stairs2003,Manchester2015,Wex2014,Kramer2017}.
However, difficulties in both the theoretical side, such as singularity and quantization problems \cite{DeWitt1967,Kiefer2007}, and the experimental side, such as mysteries of dark matter and dark energy \cite{Cline2013,Sahni2004,Debono2016,Solomon2019}, motivate people to explore innovations theoretically and search anomalies experimentally to complete our understanding of gravity. In this work, we are concerned about Lorentz symmetry whose violation is a conclusive indication of new physics beyond current standard theories \cite{AmelinoCamelia2013}.

Lorentz invariance is one of the cornerstones in both GR for gravity and the standard model for particle physics. 
In some contemporary theories, such as string theories \cite{Kostelecky1989}, loop quantum gravity \cite{Gambini1999}, brane-worlds scenarios \cite{Burgess2002}, Lorentz invariance can be broken at some energy scale. Searching for Lorentz violations is a promising way to probe unknown fundamental physics \cite{Mattingly2005}.
The theoretical framework known as the Standard-Model Extension (SME) \cite{Colladay1998,Colladay1997,Kostelecky2004} which is based on the effective field theory can characterize general violations of Lorentz invariance covering both GR and the standard model. 
Under this framework, all possible Lorentz violation terms can be constructed in the Lagrangian.

In particle sectors, intense investigations of theories and experiments on Lorentz violation have been implemented in previous works. A comprehensive summary can be found in \cite{Kostelecky2011,Tasson2014}.
Here, we focus on Lorentz violation in the gravity sector. Specifically, we consider effects induced by Lorentz violation in propagation of gravitational waves (GWs). 
The most general case of gravity sector includes couplings between gravity and matter \cite{Kostelecky2011a}. 
Since we here investigate effects of GW propagation in vacuum, we can restrict our attention to the pure gravity sector \cite{Bailey2006} and linearized level \cite{Ferrari2007}.
We also impose the gauge-invariant condition under the usual transformation $h_{\mu\nu} \rightarrow h_{\mu\nu} + \partial_\mu\xi_{\nu} + \partial_\nu\xi_{\mu}$. The discussion including gauge-violating terms be found in \cite{Kostelecky2018}.

In this gauge-invariant linearized-gravity sector of the SME, the general Lagrangian with all possible terms quadratic in metric perturbation is constructed in \cite{Kostelecky2016} where the leading-order dispersion relation for propagation of GWs is also derived. The modified dispersion relation shows that Lorentz violation can induce the effects of anisotropy, birefringence, and dispersion in propagation of GWs, which will distort observed GW signals.
The explicit waveform with the deformation is presented in \cite{Mewes2019}.
Based on this waveform, we can compare it with GW data and place constraints on coefficients characterizing Lorentz violating modification \cite{ONealAult2021}.

In past years, various experiments from microscopic to astrophysical scale focusing on Lorentz symmetry in gravity have been performed, such as, 
atom interferometry tests \cite{Mueller2008,Chung2009}, 
tests of gravitational inverse-square law at short range \cite{Shao2016,Long2015,Shao2015,BENNETT2010}, 
gravitational \v{C}erenkov radiation \cite{Schreck2018,Kostelecky2015},
gyroscopic tests \cite{Bailey2013}, 
Lunar Laser Ranging observations \cite{Bourgoin2016,Battat2007}, 
planetary orbital dynamics \cite{Iorio2012,Hees2015}, 
pulsars timing observations \cite{Shao2014a,Shao2014} and 
very long baseline interferometry \cite{PoncinLafitte2016}. 
A comprehensive review can be found in \cite{Hees2016}.

In recent years, the emerging GW astronomy provides a new tool to probe fundamental properties of gravity \cite{Yunes2016,Barack2018,Miller2019a,Schreck2017}. 
After a hundred years of being predicted, GW was first directly detected in 2015 by LIGO, Virgo, and KAGRA (LVK) Collaboration \cite{Abbott2016}. In subsequent observing runs by LVK, the detected events are accumulated at an increasing rate with continuing upgrade of detector sensitivities \cite{Abbott2016b,Abbott2019,Abbott2020,Collaboration2021e,Collaboration2021f}. The last Gravitational-Wave Transient Catalog, GWTC-3, contains 90 candidates found in observations to date \cite{Collaboration2021f}.
LVK has performed thorough tests on GR using observed GW signals including the test on Lorentz symmetry which is based on the phenomenological dispersion relation developed in \cite{Will1998,Mirshekari2012}. This approach mainly concerns about the frequency-dependent dispersion and is theory-agnostic, which can cover such as multi-fractal spacetime \cite{Calcagni2010}, massive gravity \cite{Rham2014,Will1998}, doubly special relativity \cite{AmelinoCamelia2002}, Ho\v{r}ava-Lifshitz gravity \cite{Horava2009,Vacaru2012} and theories with extra dimensions \cite{Sefiedgar2011,Vacaru2012}. 
Combined with the electromagnetic observations of binary neutron star merger GW170817 \cite{Abbott2017b,Abbott2017}, by measuring the interval between the arrival time of GWs and electromagnetic radiation, limits on Lorentz invariance considering the case of non-birefringence and non-dispersion are placed in \cite{Abbott2017a}.
Within the framework of the SME, considering the resolution of detectors to distinguish the possible split of two modes with different velocities at the frequency of maximum amplitude, the work in \cite{Kostelecky2016} present the first constraints on expansion coefficients controlling anisotropy birefringence dispersion effects with GWs.
In the works \cite{Shao2020,Wang2021f}, more events detected subsequently are taken into consideration with the same method.
However, in this method, the intervals of arrival time for the two modes are obtained by approximate estimation, which can only place constraints rather than identify deviations from GR if exist.
Birefringence effects in the propagation of GWs are also searched in GW data with both theory-agnostic approaches, such as \cite{Zhao2020} with the method concerning modes split, \cite{Wang2021g,wang2021d,Zhao2020a,Zhao2022,Wang2020} where full Bayesian analysis is perfermed but anisotropy effects are not included, and \cite{Wang2020a} considering same effects with Fisher matrix analysis, as well as model-specific approaches, such as \cite{Wu2022} considering the Nieh-Yan modified teleparallel gravity.
In the work \cite{Gong2022}, high-order spatial derivative cases are discussed.


Different from these works, within the gauge-invariant linearized gravity sector of the SME, in this article we perform full Bayesian inference with the modified waveform including anisotropy birefringence dispersion effects to investigate the Lorentz symmetry in propagation of GWs.
Our discussion is organized as follows. 
In the next section, we briefly review the gauge-invariant linearized-gravity sector of the SME and obtain the gravitational waveform based on the modified dispersion relation.
Then, in Section \ref{sec_data}, the statistical method and GW data used in this work are introduced.
We consider two cases of the lowest mass dimension $d=5,6$ which are supposed to have the most significant effects, and the results are reported in Section \ref{sec_results}.
We also present some supplementary materials in appendices including results given by individual events and the list including all events used in our analysis.
Throughout this paper, we adopt the convention where the Minkowski metric takes the signature of $(-,+,+,+)$, Greek indices $(\mu,\nu,\cdot\cdot\cdot)$ run over $0,1,2,3$ while Latin indices $(i, j, k)$ run over $1, 2, 3$, and $c=\hbar=1$.

\section{Waveform with Lorentz violation} \label{sec_theory}

In this section, we briefly review the deformation of gravitational waveform during propagation induced by the effects of Lorentz violation, which are first presented in \cite{Kostelecky2016,Mewes2019}.

\subsection{Gauge-invariant linearized-gravity sector of the SME} \label{subsec_SEM}

At the level of linearized-gravity, the metric $g_{\mu\nu}$ can be expanded in the form of $g_{\mu \nu} = \eta_{\mu\nu} + h_{\mu\nu}$ with $\eta_{\mu\nu}$ being the constant Minkowski metric.
The general quadratic Lagrangian density for the linearized-gravity sector of the SME contains all possible quadratic terms of $h_{\mu\nu}$ with arbitrary numbers of derivatives. Imposing the gauge-invariant condition under the transformation of $h_{\mu\nu} \rightarrow h_{\mu\nu} + \partial_\mu\xi_{\nu} + \partial_\nu\xi_{\mu}$, only three classes of irreducible derivative operators are left. The complete gauge-invariant quadratic Lagrangian density takes the form \cite{Kostelecky2016}
\begin{equation} \label{Lagrangian}
    \begin{aligned}
        \mathcal{L} = & \frac{1}{4}\epsilon^{\mu \rho \alpha \kappa} \epsilon^{\nu \sigma \beta \lambda} \eta_{\kappa \lambda} h_{\mu\nu} \partial_\alpha \partial_\beta h_{\rho \sigma}  \\
        &+ \frac{1}{4}h_{\mu\nu} (\hat{s}^{\mu \rho \nu \sigma} + \hat{q}^{\mu \rho \nu \sigma}+ \hat{k}^{\mu \nu \rho  \sigma}) h_{\rho \sigma},
    \end{aligned}
\end{equation}
where the first term is the usual linearized Einstein-Hilbert Lagrangian and $\epsilon^{\mu\rho \alpha \kappa}$ denotes the Levi-Civita tensor, while the second term is the three different classes of gauge-invariant operators containing the Lorentz-violating modifications.
The three operators reads
\begin{equation} \label{operators}
    \begin{aligned}
        \hat{s}^{\mu \rho \nu \sigma} =&\sum {s}^{(d) \mu \rho \alpha_{1} \nu \sigma \alpha_{2} \ldots \alpha_{d-2}} \partial_{\alpha_{1}} \ldots \partial_{\alpha_{d-2}},  \\
        \hat{q}^{\mu \rho \nu \sigma} =&\sum q^{(d) \mu \rho \alpha_{1} \nu \alpha_{2} \sigma \alpha_{3} \ldots \alpha_{d-2}} \partial_{\alpha_{1}} \ldots \partial_{\alpha_{d-2}},  \\
        \hat{k}^{\mu \nu \rho \sigma} =&\sum k^{(d) \mu \alpha_{1} \nu \alpha_{2} \rho \alpha_{3} \sigma \alpha_{4} \ldots \alpha_{d-2}} \partial_{\alpha_{1}} \ldots \partial_{\alpha_{d-2}},
    \end{aligned}
\end{equation}
where $d$ denotes the mass dimension of the three operators and the tensor coefficients control the Lorentz violation. 
The summation of $\hat{s}^{\mu \rho \nu \sigma}$ contains even numbers of $d\geq 4$. For $\hat{q}^{\mu \rho \nu \sigma}$, it contains odd numbers of $d \geq 5$, and even numbers of $d \geq 6$ for $\hat{k}^{\mu \nu \rho \sigma}$. 
The three operators have different symmetries. For $\hat{s}^{\mu \rho \nu \sigma}$, both the first pair of indices $\mu \rho$ and second pair $\nu \sigma$ are antisymmetric.
For $\hat{q}^{\mu \rho \nu \sigma}$, the first two indices are antisymmetric while the second two are symmetric.
The four indices of $\hat{k}^{\mu \nu \rho \sigma}$ are totally symmetric.
A summary of properties for the three operators can be found in Table 1 of \cite{Kostelecky2016}.

\subsection{Propagation of gravitational waves} \label{subsec_propagation}

Similar to the method in the study of photon sector of the SME \cite{Kostelecky2009}, the dispersion relation for propagation of GWs can be derived \cite{Mewes2019,Kostelecky2016} from the Lagrangian density Eq.(\ref{Lagrangian}). Assuming the Lorentz violation modification is small, the leading order dispersion relation takes the form of 
\begin{equation} \label{dispersion_relation}
    \omega = \Big(1- \zeta^0 \pm |\boldsymbol{\zeta}|\Big) |\boldsymbol{p}|,
\end{equation}
where
\begin{equation}
    \zeta^0 =  - \frac{1}{2 |\boldsymbol{p}|^2} \Big( e_{ij}^{*\rm R} \delta M^{ijmn} e^{\rm R}_{mn} + e_{ij}^{*\rm L} \delta M^{ijmn} e^{\rm L}_{mn}\Big),
\end{equation}
and
$|\boldsymbol{\zeta}|^2 = (\zeta^1)^2 + (\zeta^2)^2 + (\zeta^3)^2$ with
\begin{equation}
    \begin{aligned}
        \zeta^1 - i \zeta^2 &= \frac{1}{ |\boldsymbol{p}|^2} (e_{ij}^{*\rm R} \delta M^{ijmn} e^{\rm L}_{mn}), \\
        \zeta^1 + i \zeta^2 &= \frac{1}{ |\boldsymbol{p}|^2} (e_{ij}^{*\rm L} \delta M^{ijmn} e^{\rm R}_{mn}), \\
        \zeta^3 &= \frac{1}{2 |\boldsymbol{p}|^2} \Big( e_{ij}^{*\rm R} \delta M^{ijmn} e^{\rm R}_{mn} - e_{ij}^{*\rm L} \delta M^{ijmn} e^{\rm L}_{mn} \Big). \\
    \end{aligned}
\end{equation}
In above equations, $e^{\rm A}_{ij}$ denotes the circular polarization tensors which depends on the wave propagation direction $\boldsymbol{\hat{p}}$, satisfies the relation {$\epsilon^{i j k} \hat{p}_j e_{kl}^{\rm A} = i \rho_{\rm A} e^{i \ {\rm A}}_{\ l}$} with $\rho_{\rm R}=1$ for right-hand mode, $\rho_{\rm L} =-1$ for left-hand mode, and the normalization condition $e_{ij}^{\rm A} e_{ij}^{*\rm B} = \delta_{\rm AB}$ \cite{Takahashi2009}.
$\delta M^{ijmn}$ denotes the combination of the three classes of irreducible gauge-invariant operators
\begin{equation}
    \begin{aligned}
        \delta M^{ijmn} =& - \frac{1}{4} \left(\hat{s}^{ijmn} + \hat{s}^{ijmn}\right) - \frac{1}{2} \hat{k}^{ijmn}  \\
        &- \frac{1}{8} \left(\hat{q}^{ijmn} + \hat{q}^{ijmn} +\hat{q}^{ijmn} + \hat{q}^{ijmn}\right). \\
    \end{aligned}
\end{equation}
The above results are obtained by adopting $h_{jj}=0$, $\partial^i h_{ij} = 0$, and $h^{0 \nu} = 0$.
In the GR case, the metric perturbation $h_{\mu \nu}$ only contains two degrees of freedom. 
When the Lorentz-violating modifications are included, depending on specific types of the Lorentz violation, $h_{\mu\nu}$ may contain additional modes. The transverse and traceless conditions may not be satisfied, and the extra polarization modes of GWs exist in general. However, in the modified gravities, the amplitudes of extra polarization modes are always much smaller than the ``$+$" and ``$\times$" modes, see for instance \cite{Zhang2017a,Liu2018}. In this article, we are interested in the lead-order modification. Therefore, following  \cite{Mewes2019}, we assume additional modes are small and only consider the Lorentz violating effects in the two tensor modes.

As shown in the dispersion relation Eq.(\ref{dispersion_relation}), 
the two branches denoted by ``$\pm$'' indicate that there are two modes propagating at different phase velocities, 
\begin{equation} \label{velocities}
    v_{\pm} = 1- \zeta^0 \pm |\boldsymbol{\zeta}|.
\end{equation}
The signs ``$+$'' and ``$-$'' correspond to the fast mode $h_f$ and the slow mode $h_s$ respectively.
The fast and slow modes $(h_f, h_s)$ and the circular polarization modes $(h_{\rm R},h_{\rm L})$ are related by the normalized matrix
\begin{equation} \label{RL_fs}
    \begin{pmatrix}
        h_{\rm R} \\
        h_{\rm L} \\
    \end{pmatrix}
    =
    \begin{pmatrix}
        e^{-i \varphi/2} \cos \frac{\vartheta}{2} & -e^{-i \varphi/2} \sin \frac{\vartheta}{2} \\
		e^{i \varphi/2} \sin \frac{\vartheta}{2} & e^{i \varphi/2} \cos \frac{\vartheta}{2} \\
    \end{pmatrix}
    \begin{pmatrix}
        h_f \\
        h_s \\
    \end{pmatrix},
\end{equation}
where the angle $\vartheta$ and $\varphi$ are defined by
\begin{equation}
    \begin{aligned}
        \sin \vartheta &= \frac{\sqrt{(\zeta^1)^2+ (\zeta^2)^2}}{|\boldsymbol{\zeta}|}, \\
        \cos\vartheta &= \frac{\zeta^3}{|\boldsymbol{\zeta}|}, \\
        e^{\pm i \varphi} &= \frac{\zeta^1 \pm i \zeta^2}{\sqrt{(\zeta^1)^2+ (\zeta^2)^2}}.
    \end{aligned}
\end{equation}

The derivatives $\partial_{\mu}$ in the operator $\delta M^{ijmn}$ can be replaced with $\partial_{\mu} \rightarrow i p_{\mu}$.
Therefore, the coefficients $\zeta^0,\zeta^1,\zeta^2,\zeta^3$ are the functions of frequency $\omega$ and wave vector $\boldsymbol{p}$ of GWs. 
As discussed in \cite{Kostelecky2016,Mewes2019}, when calculating these coefficients, we can assume the GR dispersion relation $\omega =|\boldsymbol{p}| $ at the accuracy of leading order.
The direction dependent part can be decomposed by employing the spin-weighted spherical harmonics ${}_sY_{jm}$ \cite{Goldberg1967}.
The spin weight is associated with the helicity \cite{Kostelecky2009}, where $\zeta^0, \zeta^3$ have zero helicity while $\zeta^1, \zeta^2$ are helicity-4 tensors.
Therefore, the four coefficients can be decomposed as 
\begin{equation} \label{har_expansion}
    \begin{aligned}
        \zeta^0 &= \sum_{d, jm} \omega^{d-4} Y_{jm}(\boldsymbol{\hat{n}}) k_{(I)jm}^{(d)}, \\
        \zeta^1 \mp i \zeta^2 &=  \sum_{d, jm} \omega^{d-4} {}_{\pm 4}Y_{jm}(\boldsymbol{\hat{n}}) \Big[  k_{(E)jm}^{(d)} \pm i  k_{(B)jm}^{(d)}\Big], \\
        \zeta^3 &=  \sum_{d, jm} \omega^{d-4} Y_{jm}(\boldsymbol{\hat{n}}) k_{(V)jm}^{(d)},
    \end{aligned}
\end{equation}
where $\boldsymbol{\hat{n}} = -\boldsymbol{\hat{p}}$ is the direction of the GW source, and $Y_{jm}\equiv {}_0Y_{jm}$ is the usual scalar spherical harmonics. 
The indices satisfy $|s|\leq j\leq d-2$ and $m$ takes $-j, \cdots,j$.
For $m=j$ and $m=-j$, the expansion coefficients obey the relation $k^{(d)*}_{jm} = (-1)^mk^{(d)}_{j-m}$.
The mass dimension $d$ can take even numbers of $d\geq 4$ for the expansion of $\zeta^0$, odd numbers of $d\geq 5$ for $\zeta^3$, and even numbers of $d\geq 6$ for $\zeta^1 \mp i \zeta^2$. An explicit summary of these coefficients can be found in Table 1 of \cite{Mewes2019}.
The spherical harmonics are built on the Sun-centered celestial-equatorial frame following the convention \cite{Kostelecky2002,Bluhm2002,Bluhm2003}.
For convenience, the frequency and direction dependence can be separated, and we introduce a frequency-independent notation which takes the form
\begin{equation} \label{har_expansion_2}
    \begin{aligned}
        \zeta^0_{(d)} &= \sum_{jm}Y_{jm}(\boldsymbol{\hat{n}}) k_{(I)jm}^{(d)}, \\
        \zeta^1_{(d)}\mp i \zeta^2_{(d)} &=  \sum_{ jm}  {}_{\pm 4}Y_{jm}(\boldsymbol{\hat{n}}) \Big[  k_{(E)jm}^{(d)} \pm i  k_{(B)jm}^{(d)}\Big], \\
        \zeta^3_{(d)}&= \sum_{ jm} Y_{jm}(\boldsymbol{\hat{n}}) k_{(V)jm}^{(d)}.
    \end{aligned}
\end{equation}
And the phase velocity Eq.(\ref{velocities}) can be rewritten as 
\begin{equation}
    v_{\pm} = 1 - \omega^{d-4} \Big[\zeta^0_{(d)} \pm |\boldsymbol{\zeta}_{(d)}| \Big].
\end{equation}

The spherical expansion coefficients $k^{(d)}_{jm}$ are combinations of the tensor coefficients in Eq.(\ref{operators}), which equivalently control and fully characterize the Lorentz-violating modification. 
These spherical expansion coefficients are the targets which are constrained in our analysis with GW data as shown in Section \ref{sec_results}.
From the dispersion relation Eq.(\ref{dispersion_relation}) discussed in above, we can read that Lorentz-violating modification can lead to the effects of \emph{anisotropy}, \emph{dispersion}, and \emph{birefringence} in propagation of GWs, since the $\zeta_0$ and $|\boldsymbol{\zeta}|$ in Eq.(\ref{dispersion_relation}) are dependent with direction and frequency, and the two modes propagate at different velocities as shown in Eq.(\ref{velocities}).
The anisotropy effects are controlled by all expansion coefficients in Eq.(\ref{har_expansion}) except cases of $j=0$.
Excluding the coefficient $k^{(4)}_{(I)jm}$ in the expansion of $\zeta_0$, all other coefficients are relevant with the dispersion effects.
The birefringence effects are associated with $|\boldsymbol{\zeta}|$ which is governed by the expansion coefficients $k^{(d)}_{(E)jm}$, $k^{(d)}_{(B)jm}$, and $k^{(d)}_{(V)jm}$.

\subsection{Distorted waveform} \label{subsec_waveform}

In previous works \cite{Mirshekari2012,Will1998}, considering a phenomenological dispersion relation with the presence of frequency-dependent dispersion but without anisotropy and birefringence effects, the deformation of gravitational waveform during propagation have been investigated, which has been widely used in tests on GW propagation performed by LVK \cite{Collaboration2021g,Abbott2019b,Collaboration2020}. 
Following the same method, we can obtain the distorted waveform with the modified dispersion relation Eq.(\ref{dispersion_relation}) considered in this work.


Considering GW as a stream of gravitons, for two gravitons emitted at different time $t_e$ and $t'_e$ with different frequency $\omega_e$ and $\omega'_e$, the corresponding interval of arrival time $\Delta t_0$ is given by 
\begin{equation} \label{arrival_time}
    \begin{aligned}
        \Delta t_0 = & (1+z) \Delta t_e \\
        &+ \Big(\omega^{d-4}_e - \omega'^{d-4}_e\Big) \Big[-\zeta^0_{(d)} \pm |\boldsymbol{\zeta}_{(d)}| \Big] \int_{t_e}^{t_0} \frac{dt}{a^{d-3}},
    \end{aligned}
\end{equation}
where we have assumed the interval of emission time $\Delta t_e \equiv t_e -t'_e \ll a/\dot{a}$ and $z=1/a(t_e)-1$ is the redshift of GW source.
According to the derivation presented in \cite{Mirshekari2012,Will1998}, the difference of arrival time in Eq.(\ref{arrival_time}) can induce the deformation in the phase of observed GW signal. 
Assuming waveform in local wave zone of binaries is same with GR, the deformation induced during propagation takes the form
\begin{equation}
    \Psi(f) =\Psi^{\rm GR} (f) \pm \delta \Psi_1 (f, \boldsymbol{\hat{n}}) - \delta \Psi_2(f, \boldsymbol{\hat{n}}),
\end{equation}
where
\begin{equation}
    \begin{aligned}
        \delta \Psi_1(f, \boldsymbol{\hat{n}}) &= \frac{2^{d-3}}{d-3} (\pi f)^{d-3} |\boldsymbol{\zeta}_{(d)}| \int_{t_e}^{t_0} \frac{dt}{a^{d-3}}, \\
        \delta \Psi_2(f, \boldsymbol{\hat{n}})  &=  \frac{2^{d-3}}{d-3} (\pi f)^{d-3}  \zeta^0_{(d)}  \int_{t_e}^{t_0} \frac{dt}{a^{d-3}}.
    \end{aligned}
\end{equation}
The ``$\pm$'' correspond to the fast and slow modes $(h_f,h_s)$. With the phase modification, the two modes can be written as 
\begin{equation}
        h_f = h_f^{\rm GR} e^{i (\delta  \Psi_1 - \delta \Psi_2)}, \quad
        h_s = h_s^{\rm GR} e^{i (- \delta \Psi_1 - \delta \Psi_2)}.\\
\end{equation}
According to the relation shown in Eq.(\ref{RL_fs}) and the relation between the polarization modes $(h_{\rm R},h_{\rm L})$ and $(h_+,h_\times)$, we can obtain the waveform with the modification during propagation induced by Lorentz violation, 
\begin{equation} \label{waveform}
    \begin{aligned}
        h_{+} &= e^{- i \delta \Psi_2} \Big[ \big(\cos\delta \Psi_1 + i \cos\varphi \sin\vartheta \sin \delta \Psi_1\big) h_{+}^{\rm GR} \\
        &\phantom{ = e^{- i \delta \Psi_2} \Big[}
        + \big(\cos\vartheta - i \sin\vartheta \sin\varphi \big) \sin \delta \Psi_1 h_{\times}^{\rm GR}\Big], \\
        h_{\times } &= e^{- i \delta \Psi_2}  \Big[ \big(\cos\delta \Psi_1- i \cos\varphi \sin\vartheta \sin \delta \Psi_1\big)h_{\times}^{\rm GR} \\
        &\phantom{= e^{- i \delta \Psi_2}  \Big[}
        - \big(\cos\vartheta + i \sin \vartheta  \sin \varphi\big) \sin \delta \Psi_1 h_{+}^{\rm GR}\Big],
    \end{aligned}
\end{equation}
which can be directly used in parameter estimation discussed in next section.


\section{Gravitational wave data and parameter estimation} \label{sec_data}

Since the first detection of GW150914 \cite{Abbott2016}, three observing runs have been completed by LVK \cite{Abbott2016b,Abbott2019,Abbott2020,Collaboration2021e,Collaboration2021f}.
Up to now, 90 compact binary coalescence events are identified and included in the cumulative catalog, GWTC-3 \cite{Collaboration2021f}.
Among these GW events, we select 50 confident events for our analysis according to the criteria that the false alarm rate (FAR) is lower than $10^{-3}\mathrm{yr}^{-1}$ given by at least one search pipeline, which is also employed in the tests of GR performed by LVK \cite{Collaboration2021g,Abbott2019b}.
We list all these 50 events in Appendix \ref{app_events} for convenience of reference.
The strain data are downloaded from the Gravitational Wave Open Science Center\footnote{\url{https://www.gw-openscience.org/}} \cite{Abbott2021}.
Some of the events which are marked with an asterisk in Table \ref{tab_all_events} need the data after glitch subtraction for parameter estimation \cite{Abbott2020c,Cornish2015,Cornish2021a}. 
The strain data of these events are downloaded from the specific LIGO document in LIGO Document Control Center as presented in Appendix \ref{app_events}.

With the distorted waveform Eq.(\ref{waveform}), we can perform Bayesian inference with the observed GW data to estimate the probability distributions of spherical coefficients Eq.(\ref{har_expansion_2}) which can characterize Lorentz violation.
Given a waveform model $M$ described by the set of parameter $\boldsymbol{\theta}$, and background information $I$ which includes the implicit assumption for constructing the likelihood and choosing the prior, Bayes' theorem claims
\begin{equation}
p(\boldsymbol{\theta}|\boldsymbol{d}, M, I) = p(\boldsymbol{\theta}|M, I)\frac{p(\boldsymbol{d}|\boldsymbol{\theta}, M, I)}{p(\boldsymbol{d}|M, I)},
\end{equation}
where $\boldsymbol{d}$ denotes the observed data. The left-hand side is the posterior describing the probability distribution of the model parameter inferred from the observed data. 
The posterior is the consequence of two ingredients, the prior $p(\boldsymbol{\theta}|M, I)$ which implies our beliefs on the model parameters without the observation, and the likelihood $p(\boldsymbol{d}|\boldsymbol{\theta}, M, I)$ which is the probability of the realization of time series observed in the detectors under the condition of a set of specific model parameter values.
The rest term on the right-hand side is the evidence $p(\boldsymbol{d}|M, I)$ which is the normalizing factor and can also be used to compare different models.
Assuming the noise is stationary, Gaussian and uncorrelated in different detectors, the likelihood function is  \cite{Cutler1994, Romano2017,Thrane2019}
\begin{equation}
p(\boldsymbol{d}|\boldsymbol{\theta}, M, I) \propto \exp \left[-\frac12 \sum_i \left\langle \boldsymbol{h}(\boldsymbol{\theta})- \boldsymbol{d}| \boldsymbol{h}(\boldsymbol{\theta})-\boldsymbol{d} \right\rangle \right],
\end{equation}
where $\boldsymbol{h}(\boldsymbol{\theta})$ is the GW strain given by the waveform model $M$, and $i$ represents different detectors. 
The angle brackets denote the noise-weighted inner product which is defined as
\begin{equation}
\left\langle \boldsymbol{a}|\boldsymbol{b} \right\rangle = 4 \mathfrak{R} \int \frac{a(f)b^*(f)}{S_n(f)} \ {\rm d}f,
\end{equation}
where $S_n(f)$ is the noise power spectral density (PSD) of the detectors.
We use the PSD data encapsulated in LVK posterior sample releases which are expected to lead to more stable and reliable parameter estimation \cite{Abbott2019,Cornish2015,Littenberg2015}.
The spherical expansion coefficients Eq.(\ref{har_expansion_2}) of Lorentz violation are supposed to be universal for different events in the same coordinate system. 
We can directly stack the posterior of individual events to combine the information from all events,
\begin{equation} \label{combine}
p(\boldsymbol{\theta}|\{\boldsymbol{d}_i\}, M, I) \propto \prod_{i=1}^{N} p(\boldsymbol{\theta}|\boldsymbol{d}_i, M, I).
\end{equation}

To perform the parameter estimation, we employ the open-source software \texttt{Bilby} \cite{Ashton2019,RomeroShaw2020} and use the nested sampling package \texttt{Dynesty} \cite{Speagle2020}.
The waveform with Lorentz violation Eq.(\ref{waveform}) are constructed based on the GR waveform template implemented in the \texttt{LALSuite} \cite{lalsuite}. 
For BBH events, we choose the template \texttt{IMRPhenomXPHM} \cite{GarciaQuiros2020,Pratten2021,Pratten2020} which is the last generation phenomenological model including the effects of precession and higher modes.
For BNS events, we use the template \texttt{IMRPhenomD\_NRTidal} \cite{Husa2016,Khan2016,Dietrich2017} incorporating the tidal effect but neglecting the precession effect.
For the NSBH event GW200115\_042309, since the tidal effect is expected to be insignificant for the mass ratio and SNR of this detection, we employ the same BBH waveform model for this event\cite{Abbott2021a}.

\section{Results} \label{sec_results}

\begin{table} \label{tab_res_d5}
    \centering
    \begin{tabular}{cccc}
        \toprule
        $j$ & $m$ & Coefficient                 & Constraint$(10^{-15} {\rm m})$ \\
        \midrule
         0  &  0  & $k^{(5)}_{(V)00}$           & $(-1.14, 1.22)$                \\
         1  &  0  & $k^{(5)}_{(V)10}$           & $(-1.50, 1.74)$                \\
            &  1  & ${\rm Re}\ k^{(5)}_{(V)11}$ & $(-1.18, 1.02)$                \\
            &     & ${\rm Im}\ k^{(5)}_{(V)11}$ & $(-0.90, 1.10)$                \\
         2  &  0  & $k^{(5)}_{(V)20}$           & $(-6.26, 5.14)$                \\
            &  1  & ${\rm Re}\ k^{(5)}_{(V)21}$ & $(-1.66, 1.98)$                \\
            &     & ${\rm Im}\ k^{(5)}_{(V)21}$ & $(-1.50, 1.50)$                \\
            &  2  & ${\rm Re}\ k^{(5)}_{(V)22}$ & $(-2.66, 2.86)$                \\
            &     & ${\rm Im}\ k^{(5)}_{(V)22}$ & $(-1.54, 1.66)$                \\
         3  &  0  & $k^{(5)}_{(V)30}$           & $(-5.10, 4.94)$                \\
            &  1  & ${\rm Re}\ k^{(5)}_{(V)31}$ & $(-3.14, 4.10)$                \\
            &     & ${\rm Im}\ k^{(5)}_{(V)31}$ & $(-4.54, 3.86)$                \\
            &  2  & ${\rm Re}\ k^{(5)}_{(V)32}$ & $(-2.42, 2.38)$                \\
            &     & ${\rm Im}\ k^{(5)}_{(V)32}$ & $(-2.10, 1.30)$                \\
            &  3  & ${\rm Re}\ k^{(5)}_{(V)33}$ & $(-1.70, 1.74)$                \\
            &     & ${\rm Im}\ k^{(5)}_{(V)33}$ & $(-3.94, 2.54)$                \\
        \bottomrule
    \end{tabular}
    \caption{{\bf Constraints on the components in $k^{(5)}_{(V)jm}$. }  
    To place the constraints, we separately consider each component in $k^{(5)}_{(V)jm}$ and set others to be zero at one time. For each posterior point, we compute the value of the free component considered according to the values of $\zeta^3_{(5) \rm eff}$, source direction and luminosity distance of this individual  posterior point. We can obtain the posterior of the component by iterating this operation for all posterior samples. The intervals listed here are taken at $90\%$ credible level.}
\end{table}

\begin{figure}
    \centering
    \includegraphics[width=\columnwidth]{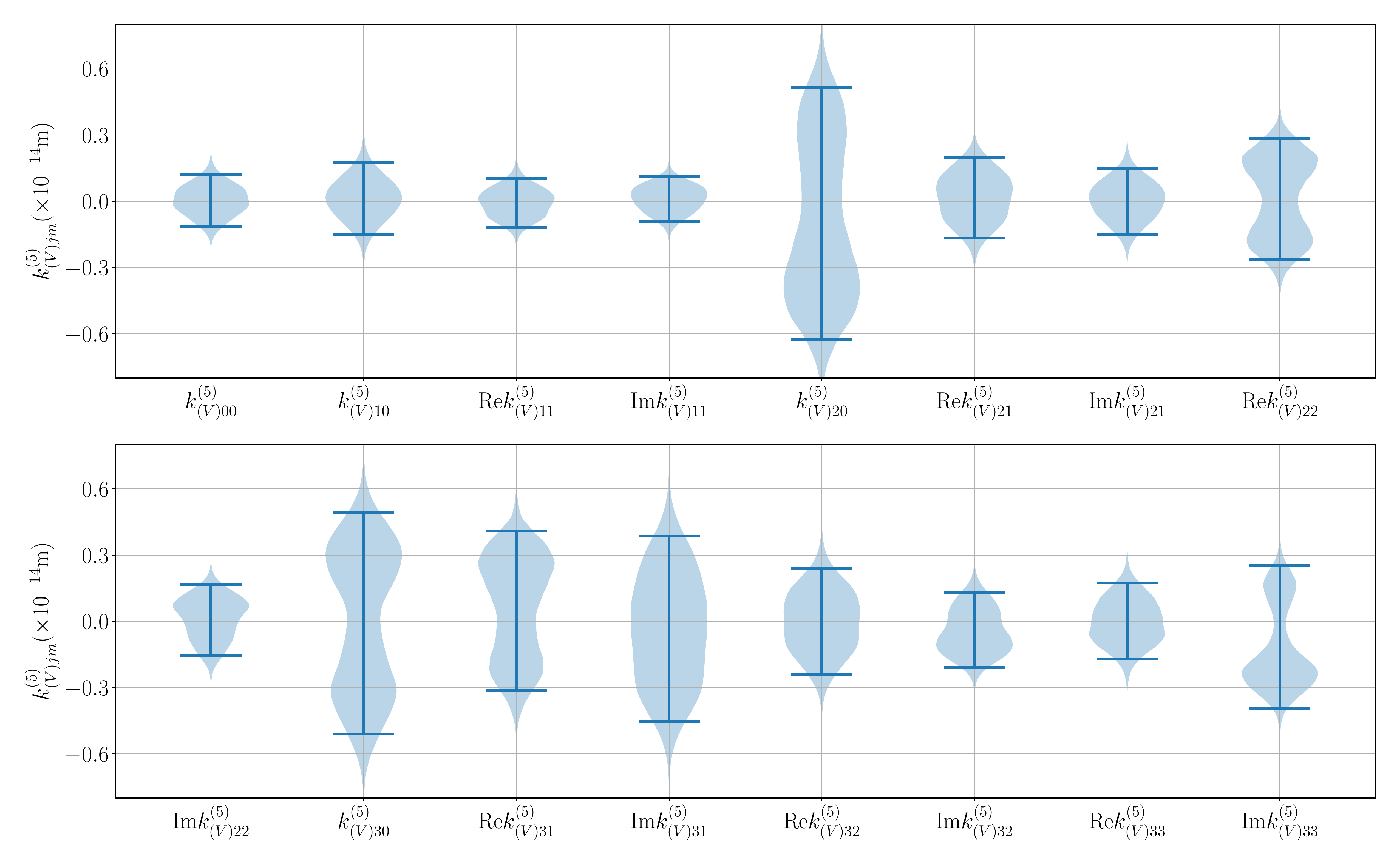}
    \caption{{\bf Combined posterior distributions of all components in $k_{(V)jm}^{(5)}$. }
    We present violin plots for posteriors of all 16 independent components in $k_{(V)jm}^{(5)}$. The error bars denote the $90\%$ credible intervals. The posterior samples all cluster around zero, and the medians or averages agree with the GR value. Although, there are bimodal features in some cases, as discussed in previous work \cite{Collaboration2020,Perkins2021,Collaboration2021f}, which could be induced by waveform systematics or data quality. These features cannot be considered as indications of deviation from GR.}
    \label{fig_d5_violin}
\end{figure}

\begin{figure}
    \centering
    \includegraphics[width=\columnwidth]{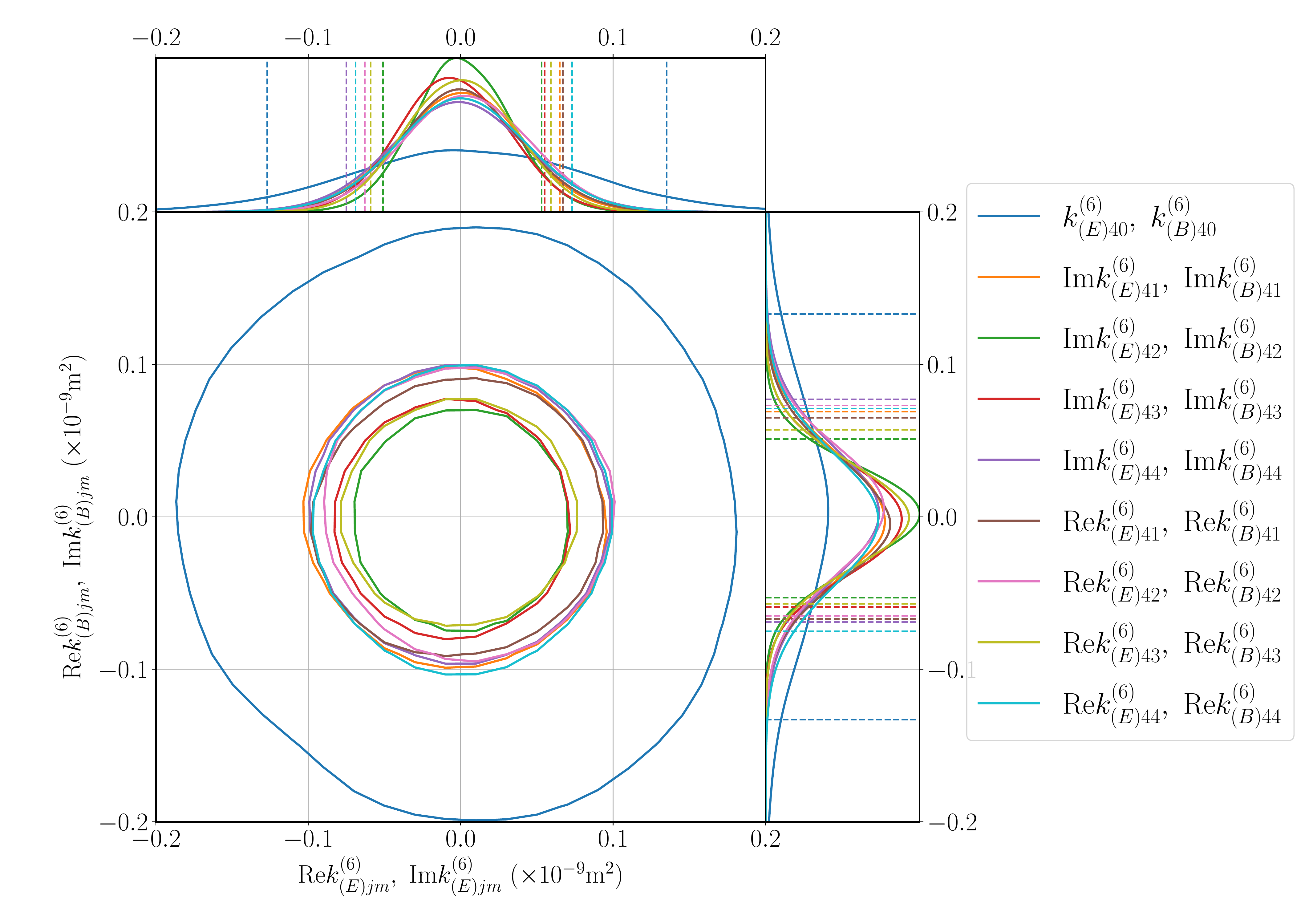}
    \caption{{\bf Combined results for the case of mass dimension $d=6$. }
    For $d=6$, expansion coefficients $k^{(6)}_{(E)jm}$ and $k^{(6)}_{(B)jm}$ have 18 free components in total. We consider a pair of them at each time. The joint distributions are shown in the main panel where we illustrate the $90\%$ credible regions. The marginalized distributions are shown in side panels with dashed lines denoting $90\%$ credible intervals.}
    \label{fig_d6_comb}
\end{figure}

{The modified dispersion relation (\ref{dispersion_relation}) indicates that during its propagation the GW can split two modes with different velocities. The previous work \cite{Kostelecky2016} consider the first GW event GW150914, take the arrival-time difference between the two modes $\Delta t \le 0.003 {\rm s}$ and a conservative value of $f\sim 100{\rm Hz}$ to obtain a heuristic constraints $\sim 10^{-14} {\rm m}$ for the case of $d=5$, and $\sim 10^{-8} {\rm m^2}$ for $d=6$. In the follow-up works \cite{Shao2020,Wang2021f}, the authors consider the accumulating GW observations, and adopt an order-of-magnitude estimation of the upper limit of the arrival-time difference $\Delta t \le 1/\rho f$. With the GW transient catalogs GWTC-1 and GWTC-2, the work \cite{Wang2021f} reports the constraints of $\sim 10^{-16} {\rm m}$ for $d=5$, and $10^{-12} {\rm m^2}$ for $d=6$.
Different from previous works which consider the possible modes split and employ the order-of-magnitude estimation, in this work, we use the distorted waveform and implement the full-Bayesian parameter estimation to obtain the constraints on the coefficients characterizing the Lorentz violation. The results are presented in this section.}

Armed with the statistical inference method discussed in the last section and the waveform derived in Section \ref{sec_theory}, we can use strain data from GW detectors to place constraints on coefficients in the expansion Eq.(\ref{har_expansion_2}) which characterize Lorentz violating modification.
The results are reported in this section.
Since terms with larger $d$ are expected to be more suppressed and induce smaller effects, interests are mainly focused on the lowest mass dimension, i.e. $d=4$ for $k^{(4)}_{(I)jm}$, $d=5$ for $k^{(5)}_{(V)jm}$, $d=6$ for $k^{(6)}_{(E)jm}$ and $k^{(6)}_{(B)jm}$.
However, as can be seen in Eq.(\ref{har_expansion}) and Eq.(\ref{dispersion_relation}), the case of $d=4$ corresponds to a frequency-independent dispersion relation. Except for an overall phase shift, there is no deformation in waveform, which means we cannot detect any effects of $k^{4}_{(I)jm}$ only with GWs. But combining with the detection of electromagnetic counterpart, the case of $d=4$ is discussed in \cite{Abbott2017a}. 
Here, we only consider the cases of $d=5$ and $d=6$ which are shown in following two subsections.

\subsection{Mass dimension $d=5$}
In the presence of Lorentz violating modification with mass dimension $d=5$, only $\zeta^3$ is non-zero. 
Assuming the correction $\delta\Psi_1$ is small and only taking the leading order, the waveform Eq.(\ref{waveform}) can be rewritten as 
\begin{equation}
    \begin{aligned}
        h_+ &= h_+^{\rm GR} + \left[ 2(\pi f)^2  \zeta^3_{(5){\rm eff}} \right] h_\times^{\rm GR},\\
        h_\times &= h_\times^{\rm GR} - \left[ 2(\pi f)^2 \zeta^3_{(5){\rm eff}} \right] h_+^{\rm GR}, \\
    \end{aligned}
\end{equation}
where $\zeta^3_{(5){\rm eff}}$ is defined as 
\begin{equation} \label{d5_para_eff}
    \zeta^3_{(5){\rm eff}} = \left[\sum_{jm} Y_{jm}(\boldsymbol{\hat{n}}) k_{(V)jm}^{(5)}\right] \int_{t_e}^{t_0} \frac{dt}{a^{2}}.
\end{equation}
$\zeta^3_{(5){\rm eff}}$ {is the parameter sampled in Bayesian inference along with other GR parameters including all intrinsic parameters characterizing the properties of GW sources and extrinsic parameters characterizing the geometric relations between sources and detectors.} The integration $\int dt/a^2$ is absorbed into $\zeta^3_{(5){\rm eff}}$ in sampling process to reduce computational costs.
The coefficients $k_{(V)jm}^{(5)}$ are reconstructed by the posterior samples of $\zeta^3_{(5){\rm eff}}$ with parameters of source location and luminosity distance.

As discussed in Section \ref{sec_theory}, for mass dimension $d=5$, the index $j$ can take $0,1,2,3$. And the index $m$ runs from $-j$ to $j$. All coefficients $k_{(V)jm}^{(5)}$ are complex numbers but need to satisfy $k^{(5)*}_{(V)jm} = (-1)^mk^{(5)}_{(V)j-m}$. 
{Hence, there are 16 free components in total.
These components are entirely tangled together. To place the constraints, when we reconstruct the probability distributions of all free components in $k^{(5)}_{(V)jm}$ from the posterior samples of $\zeta^3_{(5){\rm eff}}$, we separately consider each component and set others to be zero at one time.
{Then, we reconstruct the distribution of one component by the following process. For each posterior point of an event, we compute the value of the component according to the values of $\zeta^3_{(5){\rm eff}}$, source direction, and distance of this individual posterior point. Iterating with all posterior points, we can obtain the posterior distribution of this component.}}

{This approach is described as ``maximal-reach'' in previous works \cite{Shao2020,Wang2021f}.
Theoretically, the degeneracy of coefficients $k^{(5)}_{(V)jm}$ can be disentangled by enough events with different source locations, and we can place constraints on all free components simultaneously, if we can run Bayesian inference together with all events, which will be prohibitively slow in practice.
Appropriate approximations or assumptions are required to develop statistical methods for disentangling the degeneracy by employing GW events from different directions.
In the works \cite{Shao2020,Wang2021f}, an attempt to place global constraints is proposed, which can place limits on all components simultaneously. 
However, since we find some non-Gaussian features in the posterior of $\zeta^3_{(5){\rm eff}}$, it is not appropriate to directly follow that method in which the possible time delays are assumed to be Gaussian and the multi-Gaussian likelihood as a function of all coefficients can be constructed.
More investigations are required to allow all coefficients varying simultaneously and employ events from different directions to disentangle the degeneracy of the coefficients. We leave this in future works. We only report the results of the ``maximal-reach'' approach in this work.}

As discussed above, we reconstruct the probability density of all independent components in $k^{(5)}_{(V)jm}$ from posterior samples of $\zeta^3_{(5){\rm eff}}$ as well as source location and luminosity distance. With a fixed coordinate system, the expansion coefficients are supposed to be same for all events. Therefore, the results of individual events are combined through Eq.(\ref{combine}). The combined results are shown in Figure \ref{fig_d5_violin} and numerical values of $90\%$ credible interval are summarized in Table \ref{tab_res_d5}.
We also present results of individual events in Appendix \ref{posterior_allevents}. 
The constraints on different expansion coefficients are roughly in the same order, which means the anisotropy effect is weak even if it exists.
The probability density of most cases is clustered at zero, and the medians or averages all agree with the GR value.
But, in some cases, there are some features of bimodal and the maximum likelihood values slightly deviate from the GR value.
As discussed in previous work \cite{Collaboration2020,Perkins2021,Collaboration2021f}, these could be induced by reasons, such as waveform systematics or data quality, and cannot be seen as the indication of deviation from GR. 
As the conclusion for the case $d=5$, our results are all consistent with predictions of GR within the tolerance. We do not find any significant evidence of Lorentz violation in GW data. Thus, we place the constraints on the anisotropy birefringence dispersion coefficients $k^{(5)}_{(V)jm}$ which are in the order of $10^{-15}{\rm m}$.

\subsection{Mass dimension $d=6$}
Now we consider the case with mass dimension $d=6$. 
{Following the same treatment in previous works \citep{Kostelecky2016,Wang2021f,Shao2020} where the two cases of $d=5$ and $d=6$ are considered separately, we also discard the modification of $d=5$ when considering the case of $d=6$.}
In this case, the non-zero parameters in dispersion relation Eq.(\ref{dispersion_relation}) are $\zeta^1$ and $\zeta^2$. The waveform Eq.(\ref{waveform}) takes the form
\begin{equation} \label{d6_waveform}
    \begin{aligned}
        h_+ &= \left[1 + \frac{8i}{3} (\pi f)^3 \zeta^1_{(6){\rm eff}} \right] h_+^{\rm GR} - \left[ \frac{8i}{3} (\pi f)^3 \zeta^2_{(6){\rm eff}} \right] h_\times^{\rm GR}, \\
        h_\times &= \left[1 - \frac{8i}{3} (\pi f)^3 \zeta^1_{(6){\rm eff}} \right] h_\times^{\rm GR} - \left[ \frac{8i}{3} (\pi f)^3 \zeta^2_{(6){\rm eff}} \right] h_+^{\rm GR}, \\
    \end{aligned}
\end{equation}
where $\zeta^1_{(6){\rm eff}}$ and $\zeta^2_{(6){\rm eff}}$ are the parameters sampled in Bayesian inference.
Same with the previous case, we absorb the integration into the expansion in order to relieve the computational cost of stochastic sampling
\begin{equation} \label{d6_para_eff}
    \zeta^1_{(6){\rm eff}} \mp i \zeta^2_{(6){\rm eff}} =  \sum_{jm}  {}_{\pm 4}Y_{jm}(\boldsymbol{\hat{n}}) \Big[ k_{(E)jm}^{(6)} \pm i  k_{(B)jm}^{(6)}\Big] \int_{t_e}^{t_0} \frac{dt}{a^{3}}. \\
\end{equation}
The index $j$ can only take $4$ in this case. There are 18 independent components in $k_{(E)jm}^{(6)}$ and $k_{(B)jm}^{(6)}$ to be constrained in total.

{We employ the same method to reconstruct the probability distributions of all components in $k_{(E)jm}^{(6)}$ and $k_{(B)jm}^{(6)}$ from the posterior samples of $\zeta^1_{(6){\rm eff}}$ and $\zeta^2_{(6){\rm eff}}$ with source location and luminosity distance.
But different with the previous case, here we have two degrees of freedom in the modification introduced in the waveform.
Therefore, we choose a pair of components and set others to be zero at each time.
Same with the previous case, for every single posterior point, we convert the values of $\zeta^1_{(6){\rm eff}}$ and $\zeta^2_{(6){\rm eff}}$ to the values of the pair of components by using the source direction and distance of this posterior point.
It is natural to consider the real and imaginary parts separately, therefore, the pairs are chosen to be the real or imaginary part of $k_{(E)jm}^{(6)}$ and $k_{(B)jm}^{(6)}$ with same $j,m$ indices.
When $m=0$, the coefficients $k_{(E)40}^{(6)}$ and $k_{(B)40}^{(6)}$ are all real, and correspond to $\zeta^1_{(6){\rm eff}}$ and $\zeta^2_{(6){\rm eff}}$ respectively.
For $m \neq 0$, we reconstruct the probability distributions of expansion coefficients $k_{(E)jm}^{(6)}$ and $k_{(B)jm}^{(6)}$ from posterior samples by solving linear equations}
    \begin{equation}
        \begin{pmatrix}
            {\rm Re}\Big[ {}_{+4}Y_{jm}(\hat{\boldsymbol{n}}) + {}_{-4}Y^*_{jm}(\hat{\boldsymbol{n}}) \Big] &  - {\rm Im}\Big[ {}_{+4}Y_{jm}(\hat{\boldsymbol{n}}) + {}_{-4}Y^*_{jm}(\hat{\boldsymbol{n}}) \Big] \\   
            {\rm Im}\Big[ {}_{+4}Y_{jm}(\hat{\boldsymbol{n}}) + {}_{-4}Y^*_{jm}(\hat{\boldsymbol{n}}) \Big]  & {\rm Re}\Big[ {}_{+4}Y_{jm}(\hat{\boldsymbol{n}}) + {}_{-4}Y^*_{jm}(\hat{\boldsymbol{n}}) \Big] \\
        \end{pmatrix}
        \begin{pmatrix}
            {\rm Re}\Big[k_{(E)jm}^{(6)}\Big] \\
            {\rm Re}\Big[k_{(B)jm}^{(6)}\Big]
        \end{pmatrix}
        =
        \begin{pmatrix}
            \frac{\zeta^1_{(6){\rm eff}}}{ \int dt/a^{3}} \\
            -\frac{\zeta^2_{(6){\rm eff}}}{ \int dt/a^{3}}
        \end{pmatrix},
    \end{equation}

    \begin{equation}
        \begin{pmatrix}
            {\rm Im}\Big[ {}_{-4}Y^*_{jm}(\hat{\boldsymbol{n}}) - {}_{+4}Y_{jm}(\hat{\boldsymbol{n}}) \Big] &  {\rm Re}\Big[ {}_{-4}Y^*_{jm}(\hat{\boldsymbol{n}}) - {}_{+4}Y_{jm}(\hat{\boldsymbol{n}}) \Big] \\   
            -{\rm Re}\Big[ {}_{-4}Y^*_{jm}(\hat{\boldsymbol{n}}) - {}_{+4}Y_{jm}(\hat{\boldsymbol{n}}) \Big]  & {\rm Im}\Big[ {}_{-4}Y^*_{jm}(\hat{\boldsymbol{n}}) - {}_{+4}Y_{jm}(\hat{\boldsymbol{n}}) \Big] \\
        \end{pmatrix}
        \begin{pmatrix}
            {\rm Im}\Big[k_{(E)jm}^{(6)}\Big] \\
            {\rm Im}\Big[k_{(B)jm}^{(6)}\Big]
        \end{pmatrix}
        =
        \begin{pmatrix}
            \frac{\zeta^1_{(6){\rm eff}}}{ \int dt/a^{3}} \\
            -\frac{\zeta^2_{(6){\rm eff}}}{ \int dt/a^{3}}
        \end{pmatrix}
    \end{equation}
{for real and imaginary parts respectively.}

The combined probability distributions of all components in $k_{(E)jm}^{(6)}$ and $k_{(B)jm}^{(6)}$ are shown in Figure \ref{fig_d6_comb}.
Since we consider a pair of components at each time, we present the results by plotting the joint distributions of the two components.
In the main panel, we illustrate the $90\%$ credible regions of 2-dimension posteriors for each pairs of components in $k_{(E)jm}^{(6)}$ and $k_{(B)jm}^{(6)}$. 
As discussed previously, we consider the real and imaginary parts separately. 
And marginalized posteriors are shown in side panels. The dashed lines in side panels indicate the $90\%$ credible intervals whose numerical values are also summarized in Table \ref{tab_res_d6}.
Our results entirely support GR. We also do not find any evidence of deviation from GR in the case of $d=6$. The constraints on all components in the anisotropy birefringence dispersion coefficients $k^{(6)}_{(E)jm}$ and $k^{(6)}_{(B)jm}$ are roughly in the order of $10^{-10}{\rm m^2}$.

\begin{table*} \label{tab_res_d6}
    \centering
    \begin{tabular}{cccccc}
        \toprule
        $j$ & $m$ & $(E)$ component             & Constraint$(10^{-11} {\rm m^2})$    & $(B)$ component              & Constraint$(10^{-11} {\rm m^2})$ \\
        \midrule
         4  &  0  & $k^{(6)}_{(E)40}$           & $(-12.7, 13.5)$                   & $k^{(6)}_{(B)40}$            & $(-13.3, 13.3)$ \\
            &  1  & ${\rm Re}\ k^{(6)}_{(E)41}$   & $(-6.9, 6.7)$                     & ${\rm Re}\ k^{(6)}_{(B)41}$    & $(-6.7, 6.5)$ \\
            &     & ${\rm Im}\ k^{(6)}_{(E)41}$   & $(-6.9, 6.5)$                     & ${\rm Im}\ k^{(6)}_{(B)41}$    & $(-6.7, 6.9)$ \\
            &  2  & ${\rm Re}\ k^{(6)}_{(E)42}$   & $(-6.3, 7.3)$                     & ${\rm Re}\ k^{(6)}_{(B)42}$    & $(-6.5, 7.3)$ \\
            &     & ${\rm Im}\ k^{(6)}_{(E)42}$   & $(-5.1, 5.3)$                     & ${\rm Im}\ k^{(6)}_{(B)42}$    & $(-5.3, 5.1)$ \\
            &  3  & ${\rm Re}\ k^{(6)}_{(E)43}$   & $(-5.9, 5.9)$                     & ${\rm Re}\ k^{(6)}_{(B)43}$    & $(-5.7, 5.7)$ \\
            &     & ${\rm Im}\ k^{(6)}_{(E)43}$   & $(-6.3, 5.5)$                     & ${\rm Im}\ k^{(6)}_{(B)43}$    & $(-5.9, 5.7)$ \\
            &  4  & ${\rm Re}\ k^{(6)}_{(E)44}$   & $(-6.9, 7.3)$                     & ${\rm Re}\ k^{(6)}_{(B)44}$    & $(-7.5, 7.1)$ \\
            &     & ${\rm Im}\ k^{(6)}_{(E)44}$   & $(-7.5, 7.3)$                     & ${\rm Im}\ k^{(6)}_{(B)44}$    & $(-6.9, 7.7)$ \\
        \bottomrule
    \end{tabular}
    \caption{{\bf Bounds on all components in $k^{(6)}_{(E)jm}$ and $k^{(6)}_{(B)jm}$ at $90\%$ credible level.} Similar with $d=5$, we reconstruct posteriors of components in $k^{(6)}_{(E)jm}$ and $k^{(6)}_{(B)jm}$ after stochastic sampling process. But different with the previous case, there are two degrees of freedom in the modification. Therefore, we obtain the joint distribution of two components at each time.}
\end{table*}

\section{Summary} \label{sec_summary}

For tests on gravity, the viewpoint of effective field theory is a compromise between tests on specific models, such as \cite{Perkins2021,Niu2021}, and the most general model-independent tests of GR, such as the residual test performed by LVK \cite{Collaboration2021g,Abbott2019b,Collaboration2020}, which can take advantages of both.
Constraints given by the approach of effective field theory can usually have explicit physical meanings, in the meanwhile can cover different theories simultaneously \cite{Zhao2020a,Wang2021g,wang2021d,Qiao2019}.
In this work, we consider a piece of the effective field theory, the gauge-invariant linearized-gravity sector of the SME, to investigate the Lorentz symmetry during propagation of GWs.

The waveform with Lorentz violating modification is reviewed in Section \ref{sec_theory}.
We focus on propagation effects, while generation of GWs is assumed to be consistent with GR. We start by the Lagrangian in the gauge-invariant linearized-gravity sector in the SME \cite{Kostelecky2016} which contains all possible gauge-invariant quadratic terms of the metric perturbation $h_{\mu\nu}$ and can characterize general Lorentz violating modification. 
Following the similar method developed in the discussion of Lorentz symmetry in the photon sector of the SME \cite{Kostelecky2009}, the modified dispersion relation for propagation of GWs can be obtained from the Lagrangian.
As can be read from this dispersion relation, GWs can split two propagation modes with different velocities. The velocities are dependent with the direction and frequency of GWs. The anisotropy, birefringence, and dispersion effects are presented in propagation of GWs due to the Lorentz violating modification.
Based on the modified dispersion relation, the deformation of waveform including the Lorentz violating effects can be derived \cite{Mewes2019}, which allows us to perform full Bayesian analysis with GW data to constrain these effects.

In the last GWs events catalog, GWTC-3, 90 candidates are included. We select 50 confident events by referring to the tests on GR performed by LVK \cite{Collaboration2021g,Abbott2019b} in our analysis.
With the waveform including the deformation induced by Lorentz violation, we perform full Bayesian analysis to place the constraints on the coefficients characterizing the anisotropy, birefringence, and dispersion effects. Results are presented in Section \ref{sec_results}. 

Since the higher mass dimension are supposed to be more suppressed and there are no detectable effects only with GWs for the case of $d=4$, we focus on the cases of $d=5$ and $d=6$. 
Due to the large number of free components in the expansion coefficients $k_{(V)jm}^{(5)}$, $k_{(E)jm}^{(6)}$ and $k_{(B)jm}^{(6)}$, we consider the entirety of them and sample the effective parameters as shown by Eq.(\ref{d5_para_eff}) and Eq.(\ref{d6_para_eff}) in the process of stochastic sampling. The posteriors of all independent components are reconstructed from the samples of the effective parameters with source location and luminosity distance. 
In the reconstruction, we employ the approach that is called  ``maximal-reach'' in previous works \cite{Shao2020,Wang2021f}. We consider one component in $k_{(V)jm}^{(5)}$ or a pair of components in $k_{(E)jm}^{(6)}$ and $k_{(B)jm}^{(6)}$ at each time, and set others to be zero. 

Our results are consistent with GR, we do not find any evidence for Lorentz violation in GW data.
Thus, we can present constraints on the anisotropy birefringence dispersion coefficients.
The $90\%$ limits on components in $k_{(V)jm}^{(5)}$ are roughly in the order of $10^{-15}{\rm m}$, for components in $k_{(E)jm}^{(6)}$ and $k_{(B)jm}^{(6)}$, the $90\%$ bounds are in the order of $10^{-10}{\rm m^2}$. Our $90\%$ bounds are corresponding with results presented in previous works \cite{Shao2020,Wang2021f} which obtain the constraints by considering the resolution ability of detectors to distinguish the possible split of GWs induced by two propagation modes.
Although, the constraints reported here are slightly weaker than the previous bounds \cite{Shao2020,Wang2021f}, 
the constraints are obtained by full Bayesian analysis here whereas the previous bounds are based on the approximate estimation for intervals of arrival time of two modes, which can only place constraints but cannot identify deviation from GR if exist. 


\acknowledgments

Tao Zhu appreciates the helpful discussions with Bo Wang, Sen Yang, Qiang Wu, and Lijing Shao. This work is supported by the National Key Research and Development Program of China Grants No. 2021YFC2203102, 2022YFC2200100 and 2020YFC2201503, the Zhejiang Provincial Natural Science Foundation of China under Grants No. LR21A050001 and No. LY20A050002, the NSFC grants No.12273035, 11633001, 11653002, 11603020, 11903030, 12003029, 12275238, 11903033, the Fundamental Research Funds for the Central Universities under Grants No. WK2030000036, WK3440000004 and WK2030000044, the Strategic Priority Research Program of the Chinese Academy of Sciences Grant No. XDB23010200, and the China Manned Space Program through its Space Application System, and the China Postdoctoral Science Foundation grant No.2019M662168.

This research has made use of data or software obtained from the Gravitational Wave Open Science Center (gw-openscience.org), a service of LIGO Laboratory, the LIGO Scientific Collaboration, the Virgo Collaboration, and KAGRA. LIGO Laboratory and Advanced LIGO are funded by the United States National Science Foundation (NSF) as well as the Science and Technology Facilities Council (STFC) of the United Kingdom, the Max-Planck-Society (MPS), and the State of Niedersachsen/Germany for support of the construction of Advanced LIGO and construction and operation of the GEO600 detector. Additional support for Advanced LIGO was provided by the Australian Research Council. Virgo is funded, through the European Gravitational Observatory (EGO), by the French Centre National de Recherche Scientifique (CNRS), the Italian Istituto Nazionale di Fisica Nucleare (INFN) and the Dutch Nikhef, with contributions by institutions from Belgium, Germany, Greece, Hungary, Ireland, Japan, Monaco, Poland, Portugal, Spain. The construction and operation of KAGRA are funded by Ministry of Education, Culture, Sports, Science and Technology (MEXT), and Japan Society for the Promotion of Science (JSPS), National Research Foundation (NRF) and Ministry of Science and ICT (MSIT) in Korea, Academia Sinica (AS) and the Ministry of Science and Technology (MoST) in Taiwan.

Data analyses and results visualization in this work made use of \texttt{Bilby}\cite{Ashton2019}, \texttt{Dynesty}\cite{Speagle2020}, \texttt{LALSuite}\cite{lalsuite}, \texttt{PESummary}\cite{Hoy2021a}, \texttt{NumPy}\cite{Harris2020, Walt2011}, \texttt{SciPy}\cite{Virtanen2020} and \texttt{matplotlib}\cite{Hunter2007}.

\bibliography{ref}
\bibliographystyle{JHEP}

\newpage

\appendix

\section{Posteriors for individual events} \label{posterior_allevents}

\begin{figure}[p]
    \centering
    \includegraphics[width=0.45\columnwidth]{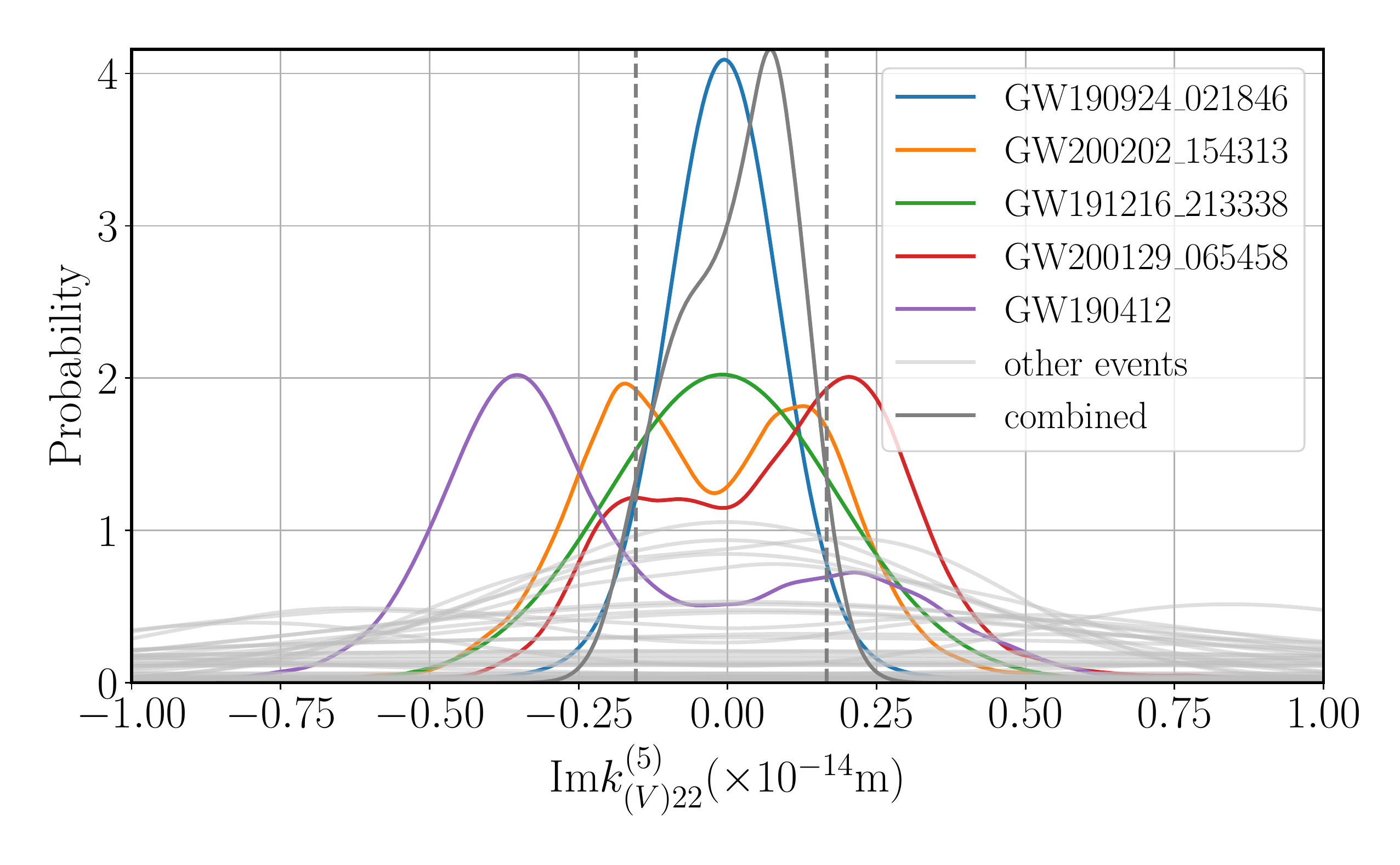}
    \includegraphics[width=0.45\columnwidth]{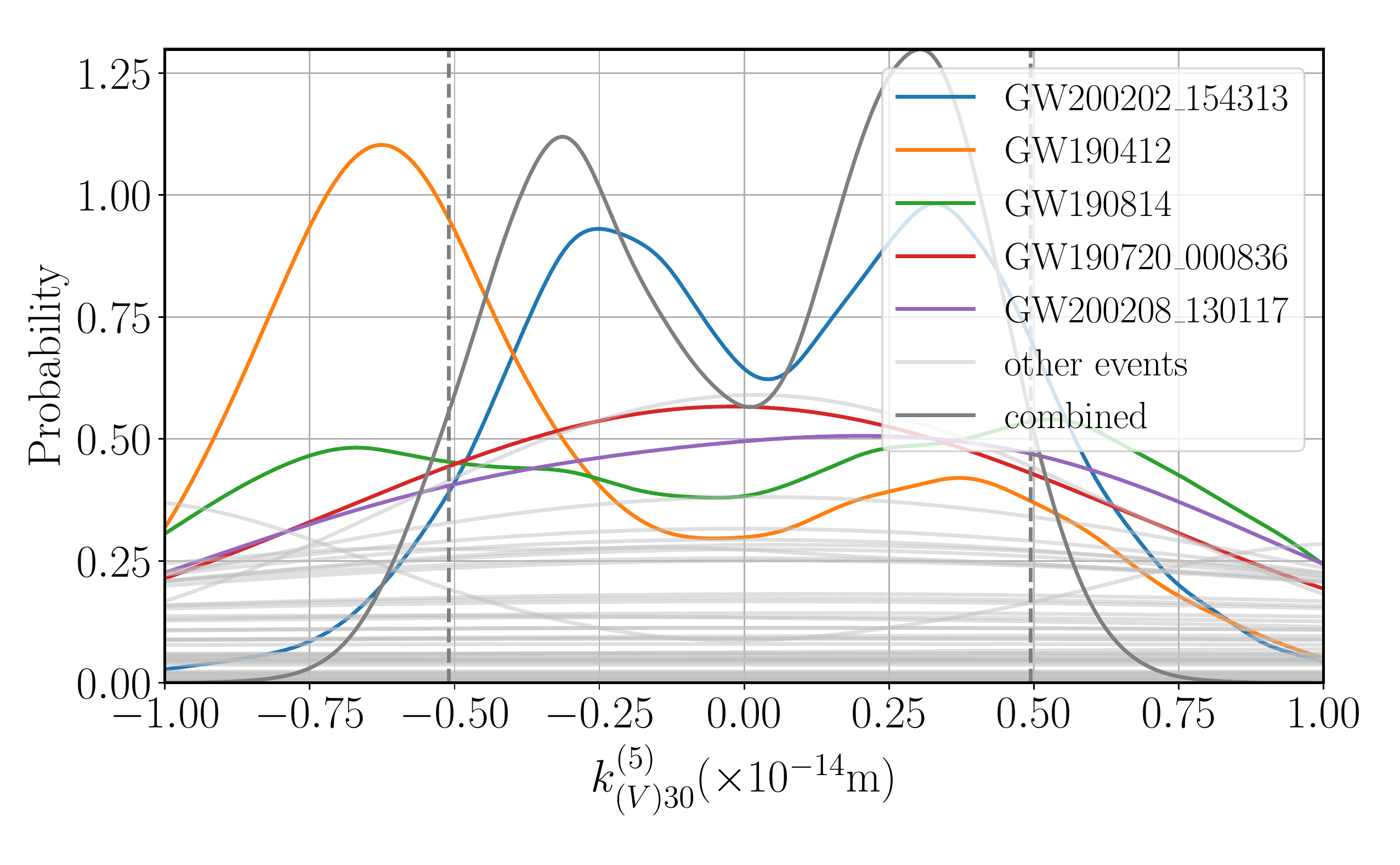}
    \includegraphics[width=0.45\columnwidth]{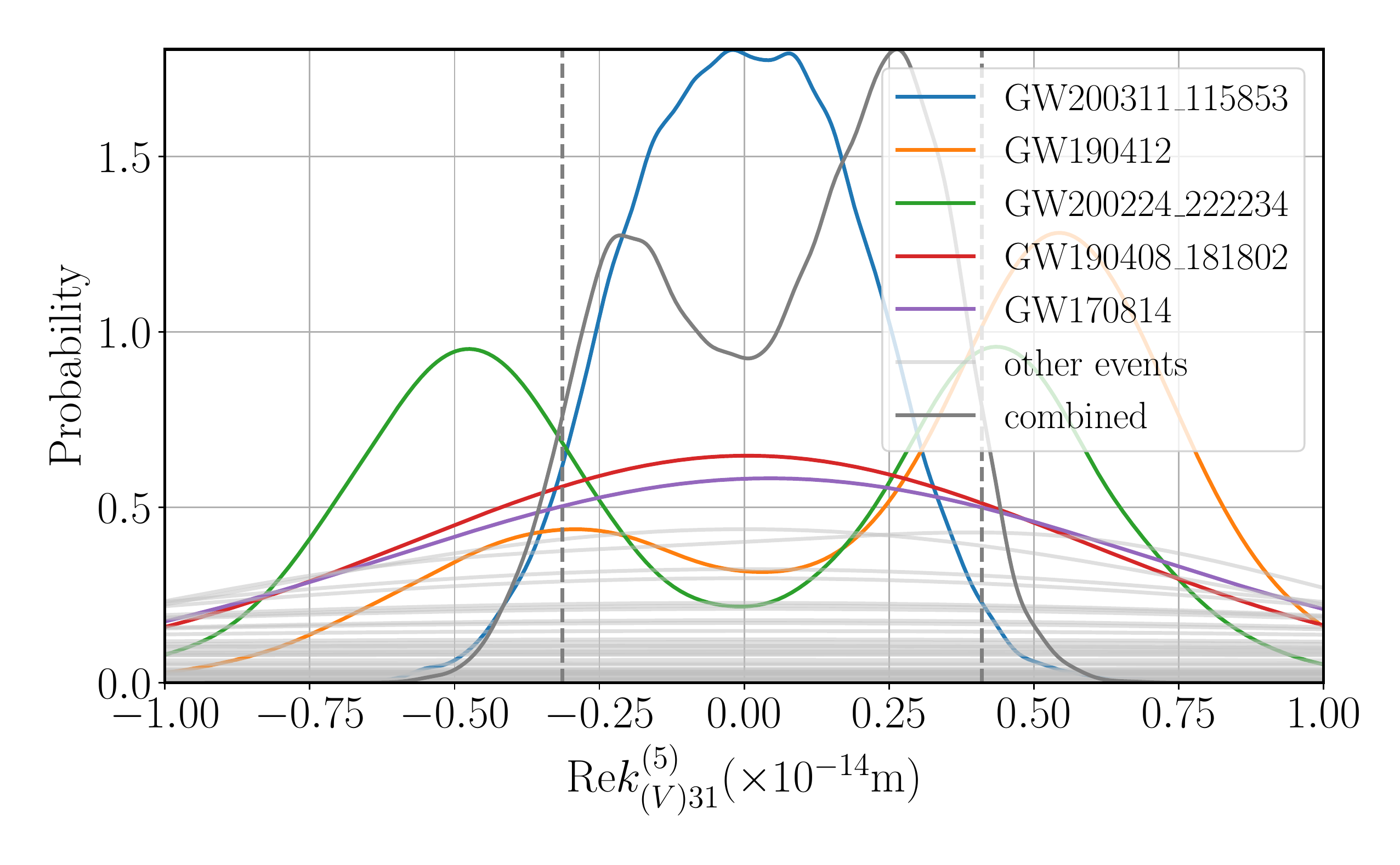}
    \includegraphics[width=0.45\columnwidth]{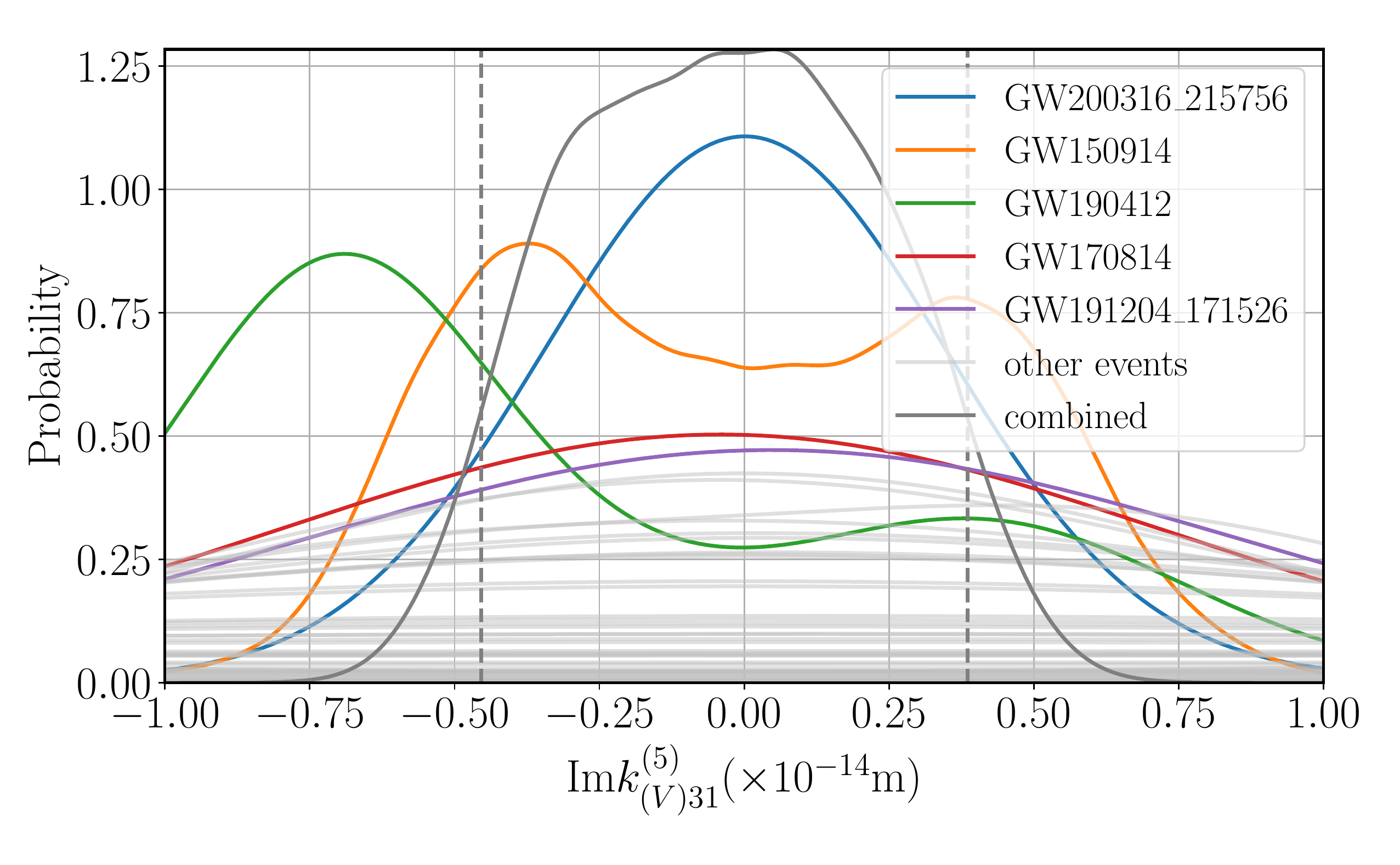}
    \includegraphics[width=0.45\columnwidth]{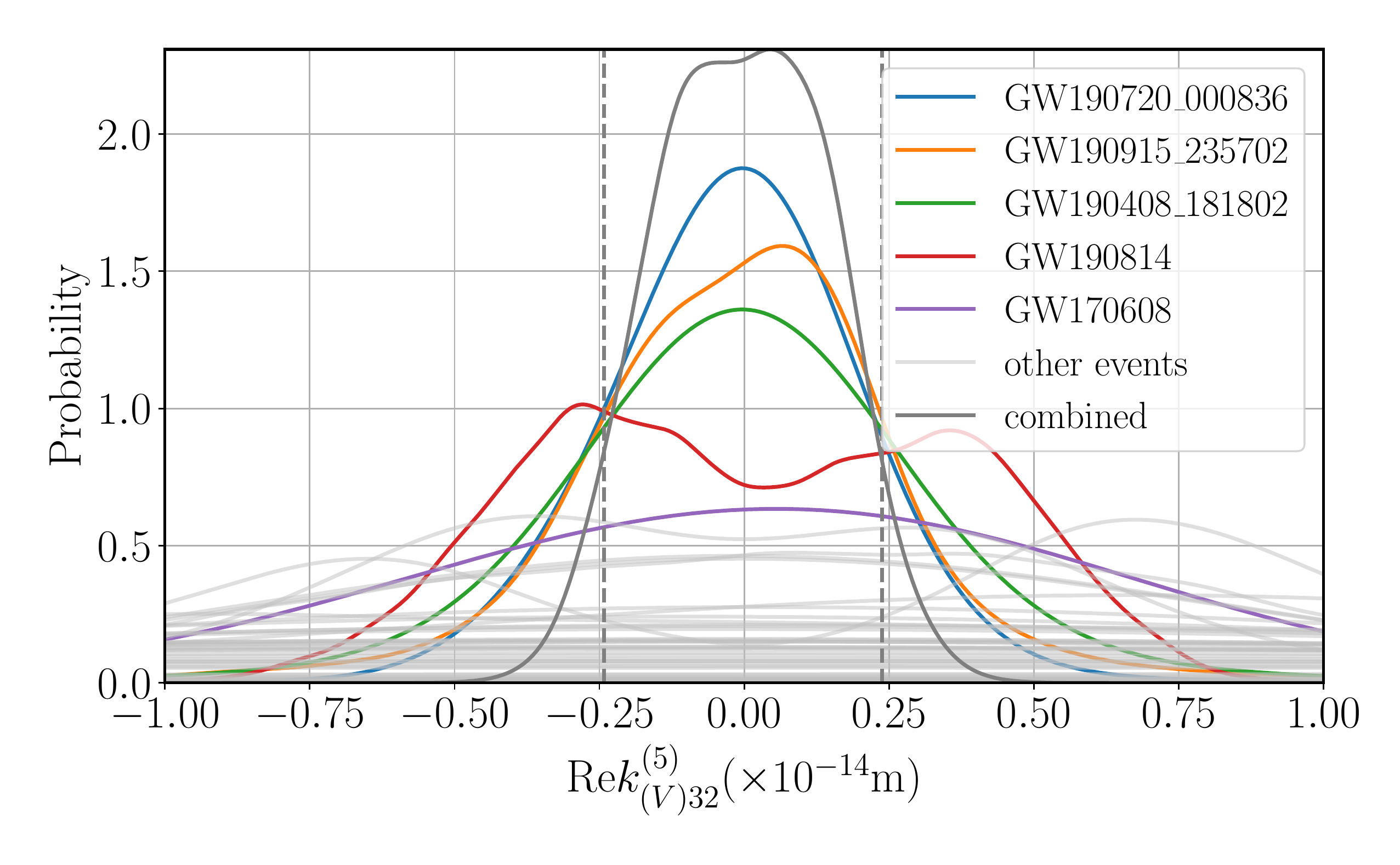}
    \includegraphics[width=0.45\columnwidth]{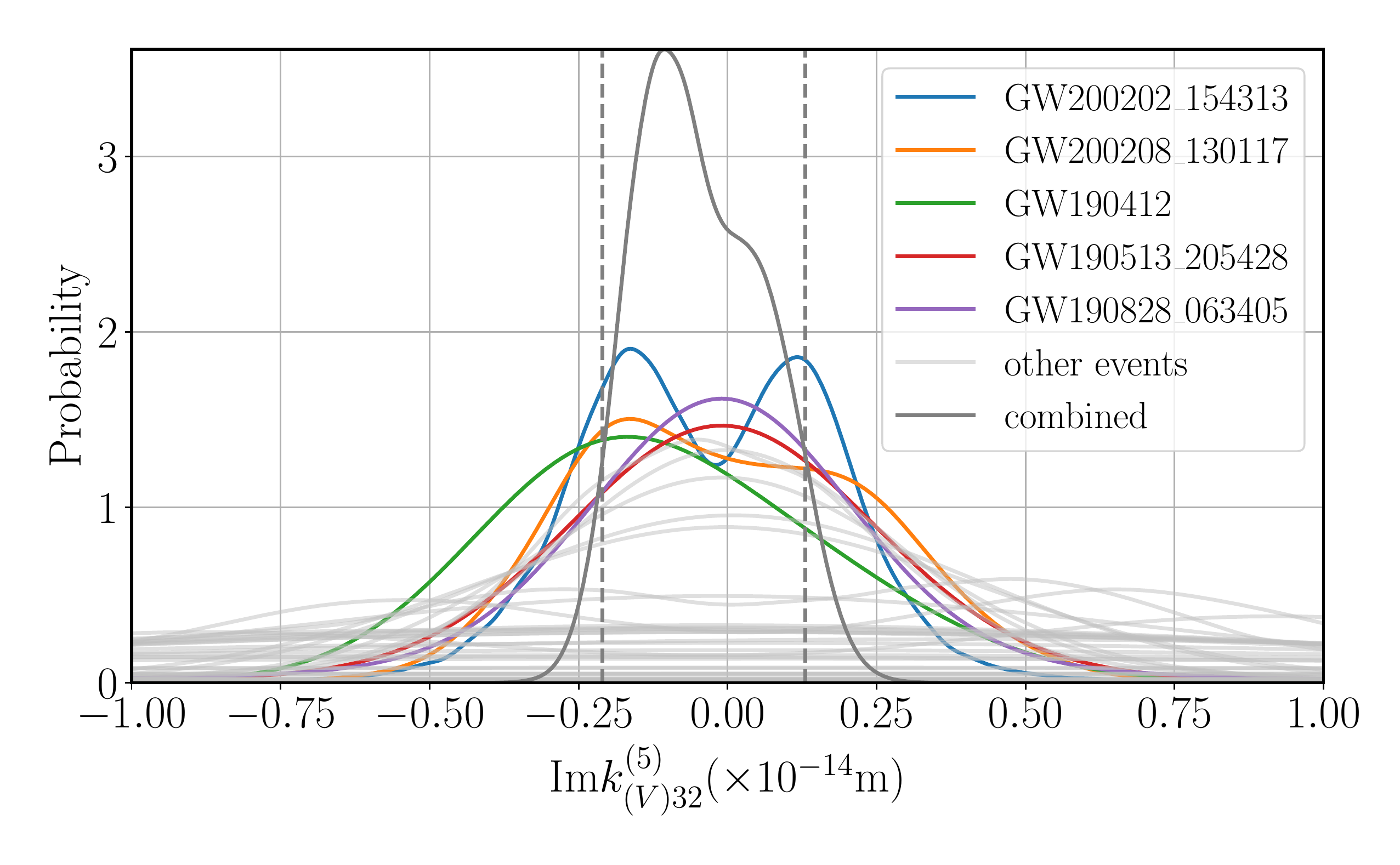}
    \includegraphics[width=0.45\columnwidth]{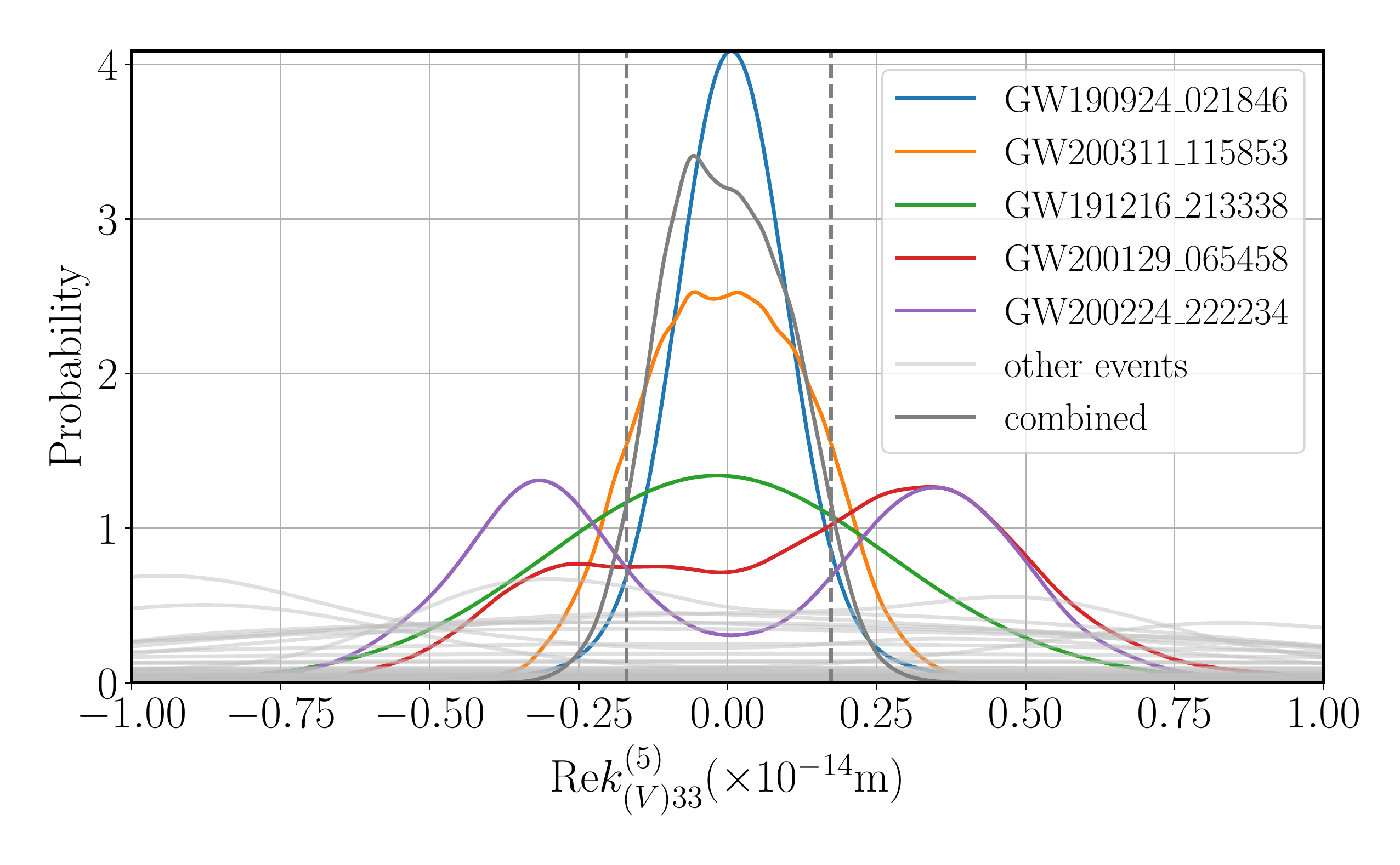}
    \includegraphics[width=0.45\columnwidth]{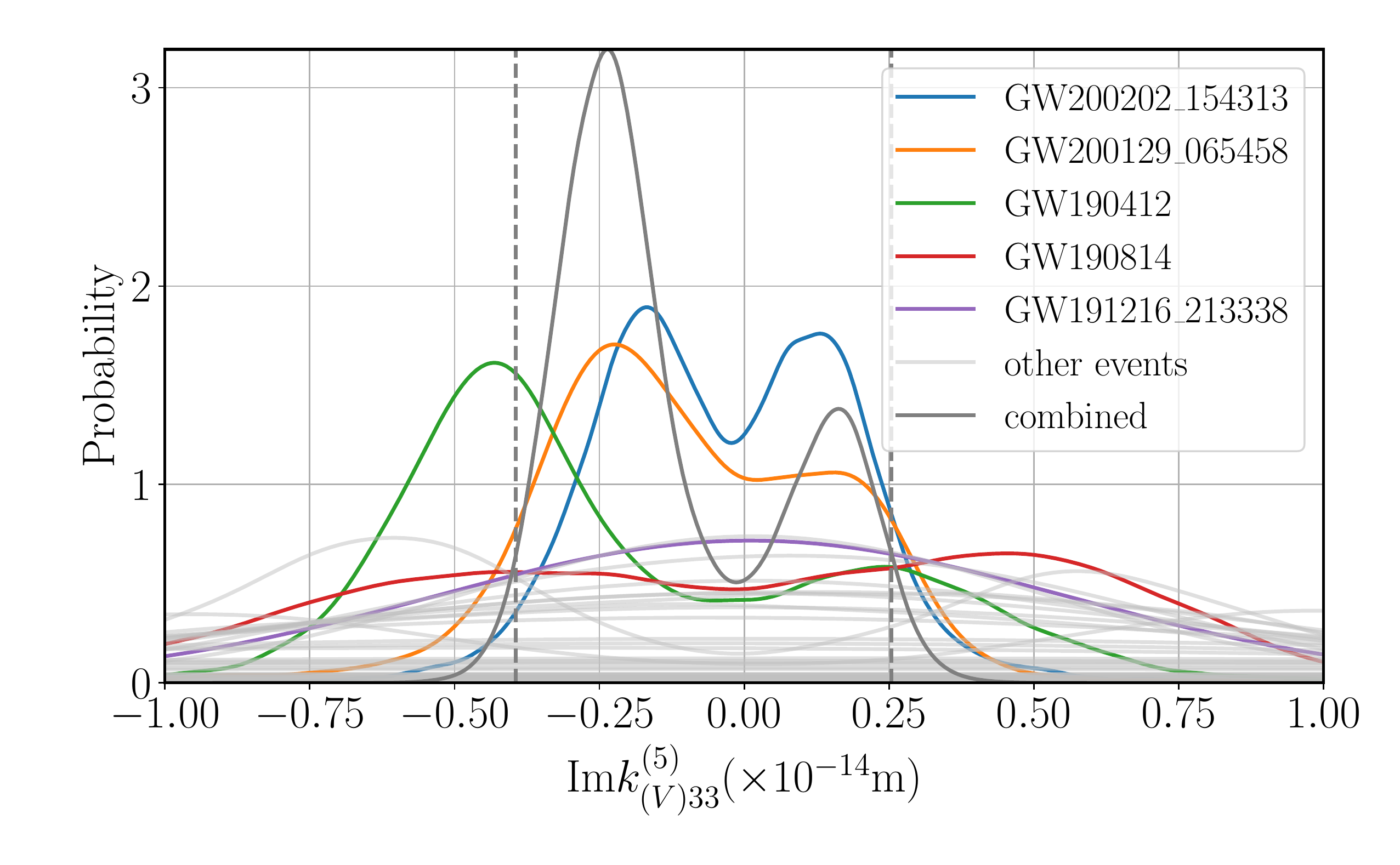}
    \caption{{\bf Posteriors of 16 components in $k_{(V)jm}^{(5)}$ for individual events. } 
    The events with the tightest constraints are highlighted by colors, while others are illustrated by light silver. The combined results are presented by gray solid lines with the dashed lines denoting $90\%$ credible intervals.}
    \label{fig_app_d5}
\end{figure}

\begin{figure}[p]
    \centering
    \includegraphics[width=0.45\columnwidth]{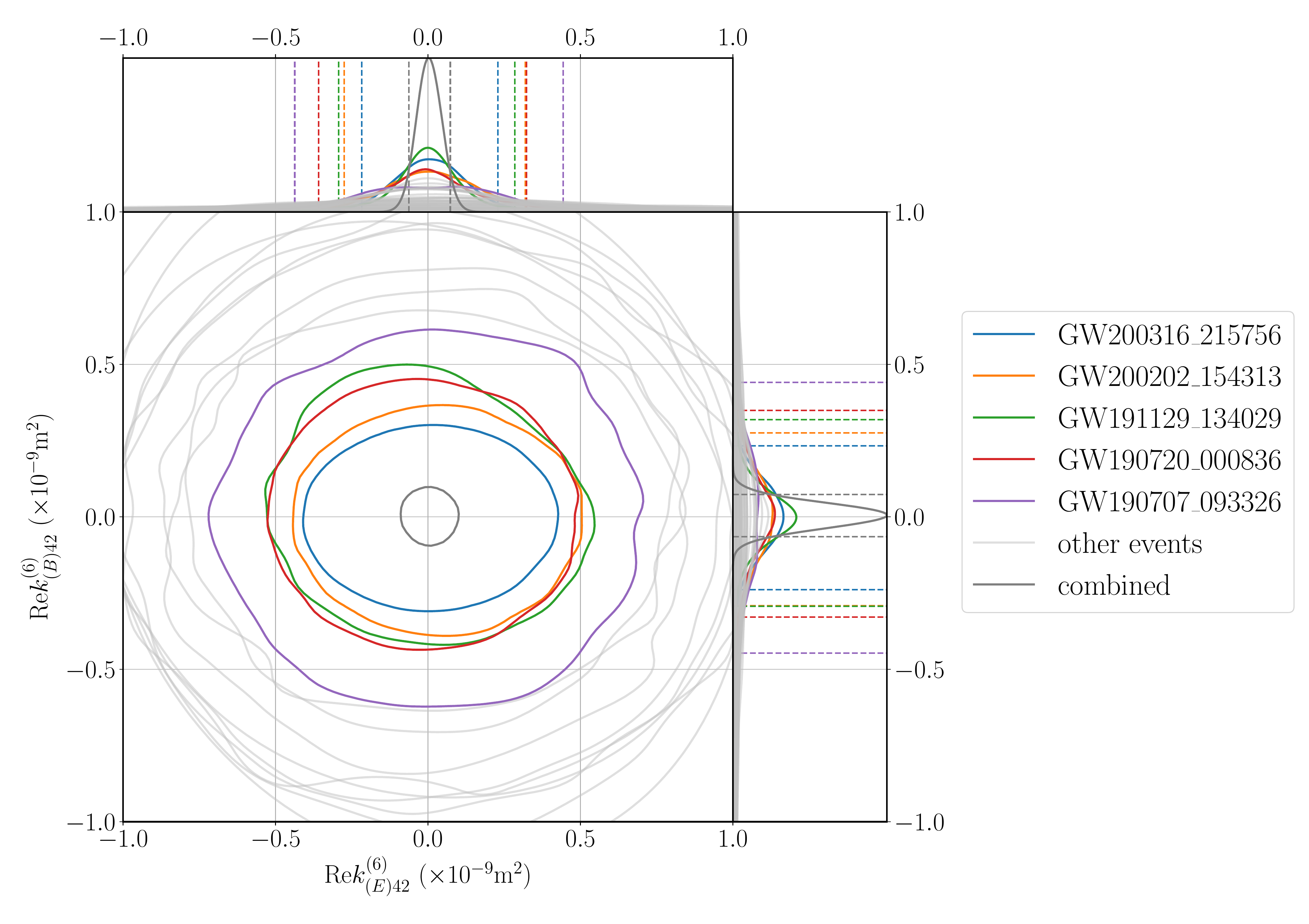}
    \includegraphics[width=0.45\columnwidth]{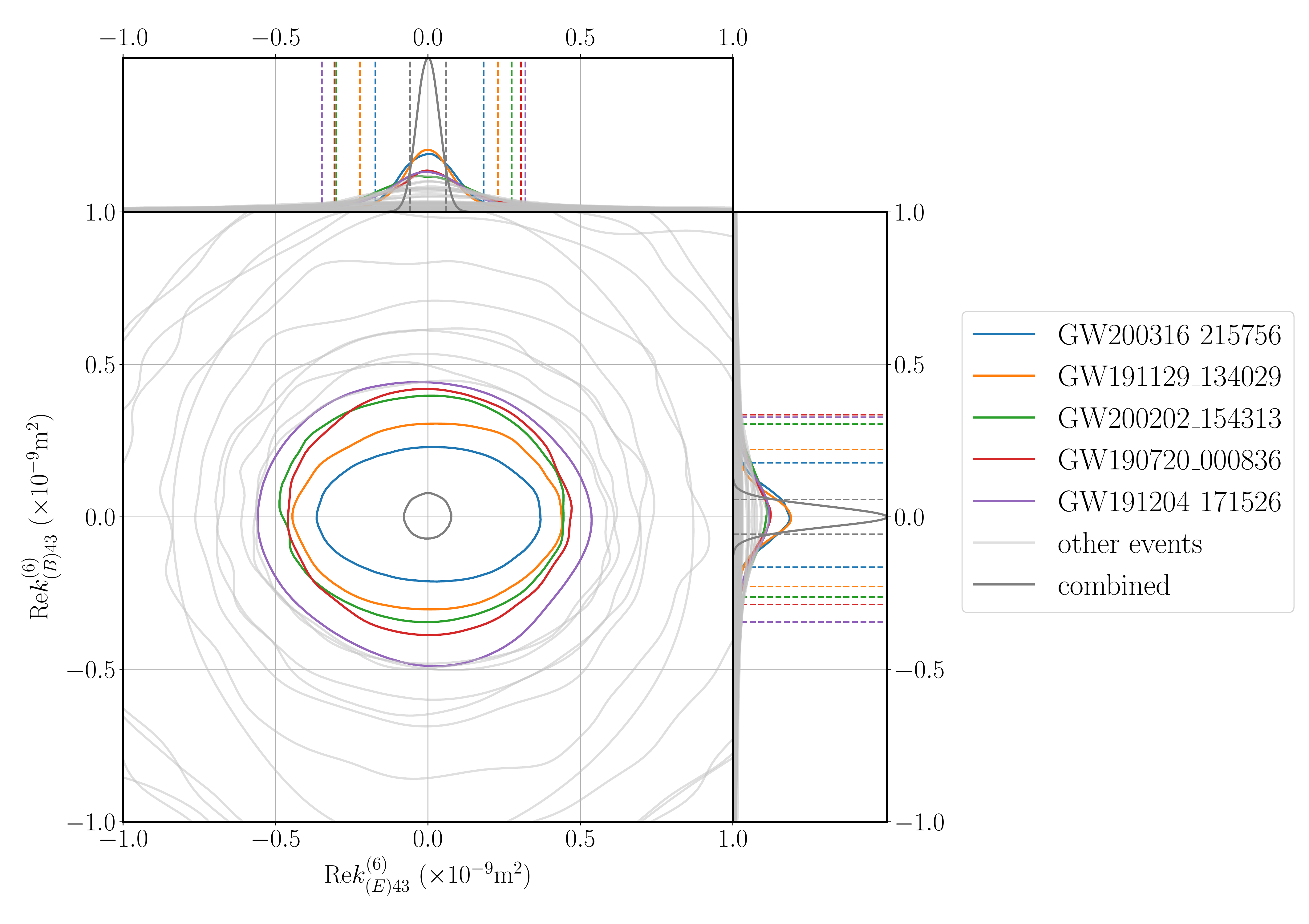}
    \includegraphics[width=0.45\columnwidth]{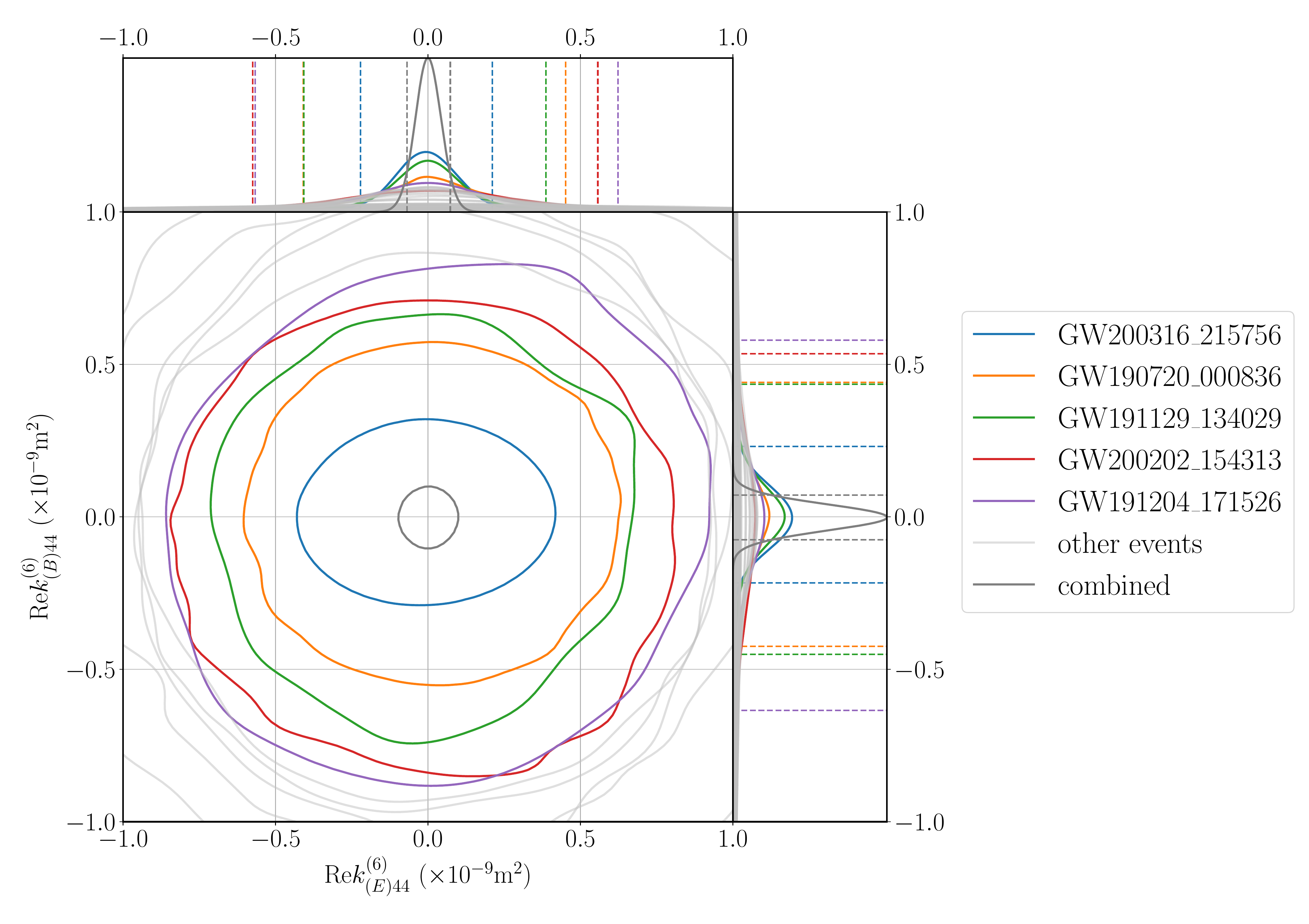}
    \caption{{\bf Probability distributions of all components in $k^{(6)}_{(E)jm}$ and $k^{(6)}_{(B)jm}$. }
    For the case of $d=6$, there are two degrees of freedom in the modification of waveform as shown in Eq.(\ref{d6_waveform}). When we reconstruct the posteriors of all components in $k^{(6)}_{(E)jm}$ and $k^{(6)}_{(B)jm}$., we consider a pair of them at each time. The results are shown by the joint distribution. 
    The $90\%$ credible regions are plotted in main panels while the marginalized distributions are presented in side panels where dashed lines are used to denote $90\%$ credible intervals.
    Same with the previous case, the 5 events whose $90\%$ credible regions have the smallest area are highlighted by colors.
    The combined results shown in main text are also illustrated here by gray color for convenience of comparison.}
    \label{fig_app_d6}
\end{figure}

\begin{figure}
    \centering
    \includegraphics[width=0.45\columnwidth]{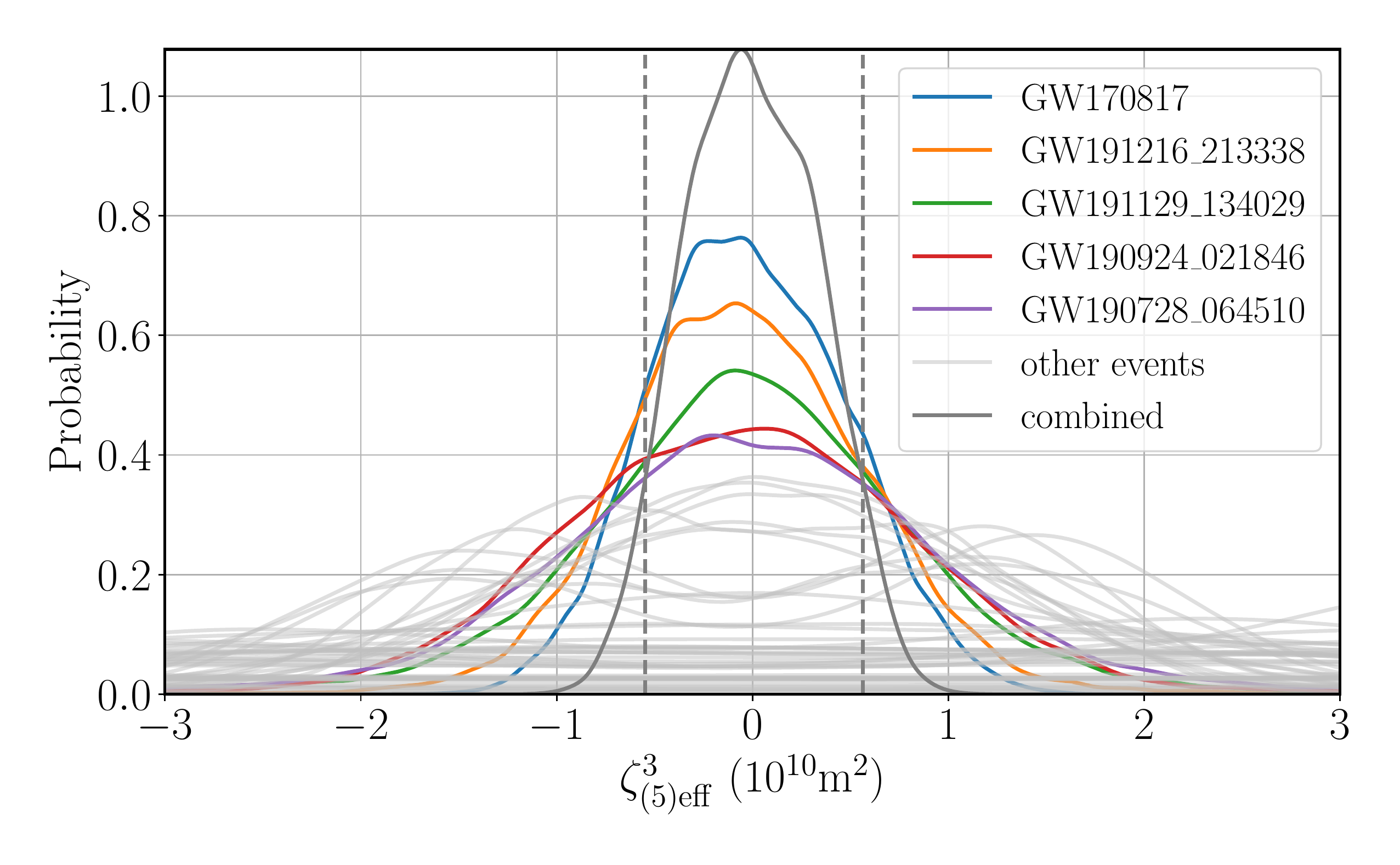}
    \caption{{\bf  Posterior distributions of $\zeta^3_{(5){\rm eff}}$.}
    Similar to Figure \ref{fig_app_d5}, the posteriors of $\zeta^3_{(5){\rm eff}}$ which is the parameter sampled in the stochastic sampling process are presented in this figure with colored lines highlighting the 5 events giving the tightest constraints. The dashed gray lines denote the $90\%$ credible intervals of combined results.}
    \label{fig_d5_zeta3}
\end{figure}

\begin{figure}
    \centering
    \includegraphics[width=0.45\columnwidth]{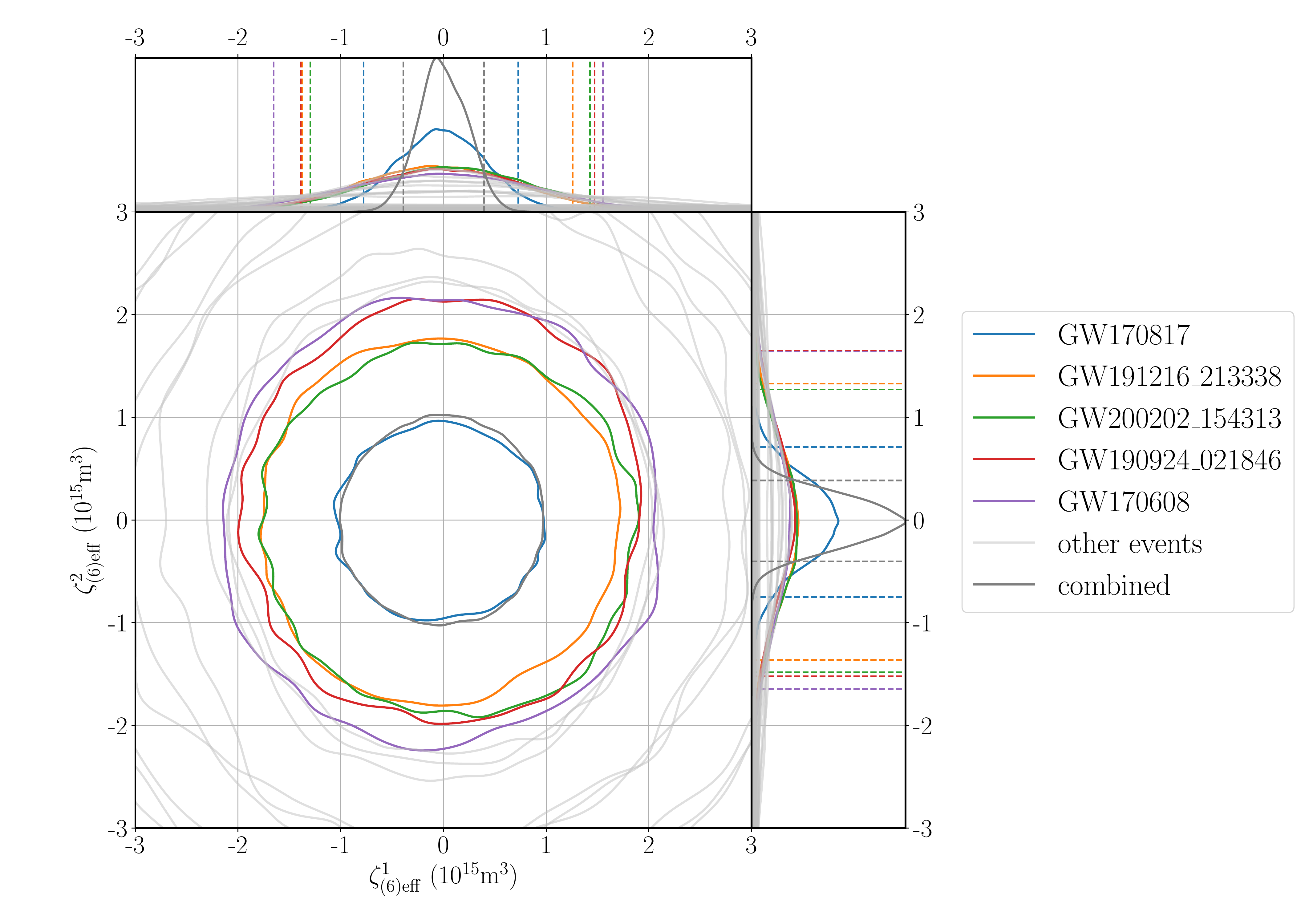}
    \caption{{\bf Joint posterior probability distributions of $\zeta^1_{(6){\rm eff}}$ and $\zeta^2_{(6){\rm eff}}$.}
    Same with Figure \ref{fig_app_d6}, the main panel shows the $90\%$ credible regions where the colored lines represent the 5 events whose $90\%$ credible regions have smallest area. The marginalized distributions are presented in the side panel with dashed lines denoting $90\%$ credible intervals.}.
    \label{fig_d6_zeta1zeta2}
\end{figure}

We present all results given by individual events here. The posteriors for the case of $d=5$ are shown in Figure \ref{fig_app_d5}, and the case of $d=6$ are in Figure \ref{fig_app_d6}. Each subfigure represents one component in expansion coefficients $k_{(V)jm}^{(5)}$ or a pair of components in $k^{(6)}_{(E)jm}$ and $k^{(6)}_{(B)jm}$. Results of all events considered in this work are plotted in each subfigure.
However, in order to avoid confusion, we only highlight the 5 events which can provide the tightest constraints by colors, while others are denoted by light silver. 
For the case of $d=6$, due to the limitation of view size, not all events are presented in main panels.
The combined results are also shown in these figures by solid lines with dashed lines denoting $90\%$ credible intervals.
{Similar to Figure \ref{fig_app_d5} and Figure \ref{fig_app_d6}, the posteriors of parameters $\zeta^3_{(5){\rm eff}}$, $\zeta^1_{(6){\rm eff}}$ and $\zeta^2_{(6){\rm eff}}$ which are directly sampled in the stochastic sampling process are also shown in Figure \ref{fig_d5_zeta3} and Figure \ref{fig_d6_zeta1zeta2} respectively. }

\section{Events used in the analysis} \label{app_events}

The last GW events catalog, GWTC-3, contains 90 candidates found in the three observing runs up to date \cite{Collaboration2021f}. Among them, we select 50 confident events which are listed in Table \ref{tab_all_events} by the criterion that at least in one search pipeline the false alarm rate is lower than $10^{-3}\mathrm{yr}^{-1}$.
This criterion is chosen by referring to the tests on GR performed by LVK \cite{Collaboration2021g,Abbott2019b}. 
The strain data of these events are downloaded from \href{https://www.gw-openscience.org}{Gravitational Wave Open Science Center}.

But some of them are contaminated by glitches which may cause poor convergence and bias in parameter estimation as discussed in \cite{Powell2018,Chatziioannou2021a}.
The reconstruction and subtraction of glitches have been performed by LVK with the BayesWave algorithm \cite{Littenberg2016,Cornish2015, Cornish2021a}. The cleaned data can be downloaded from LIGO Document Control Center separately. 
For the events, including
GW170817,
GW190425,
GW190503\_185404,
GW190513\_205428,
GW190924\_021846,
GW191109\_010717,
GW200115\_042309 and 
GW200129\_065458,
we use the data after glitch-subtraction downloaded from 
(\url{https://dcc.ligo.org/LIGO-T1700406/public}),
(\url{https://dcc.ligo.org/LIGO-T2000470/public}),
(\url{https://dcc.ligo.org/T1900685/public}) and
(\url{https://zenodo.org/record/5546680#.YanbUvFBy8o})
in the procedure of parameter estimation.

\begin{figure}[p]
    \centering
    \includegraphics[width=0.45\columnwidth]{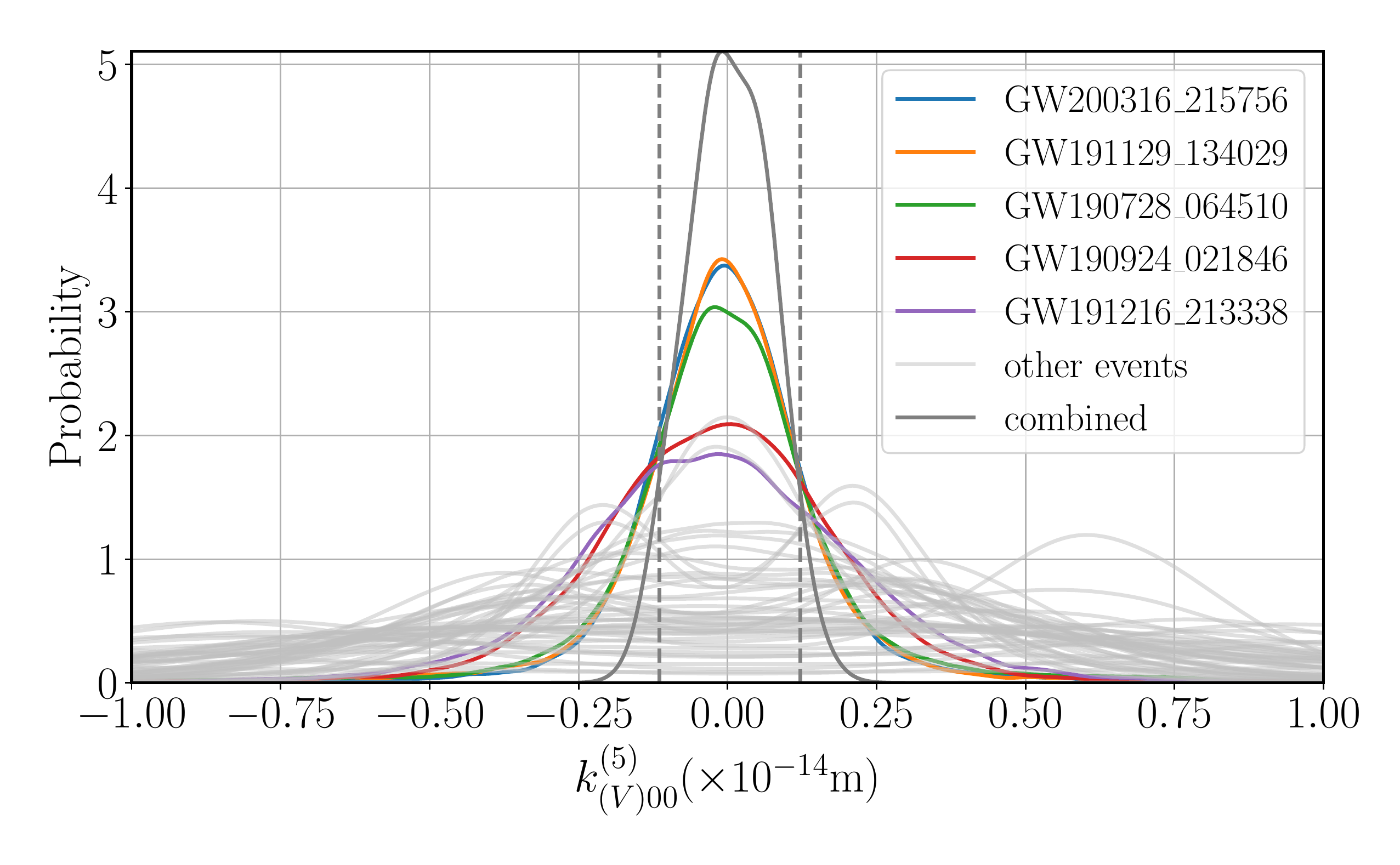}
    \includegraphics[width=0.45\columnwidth]{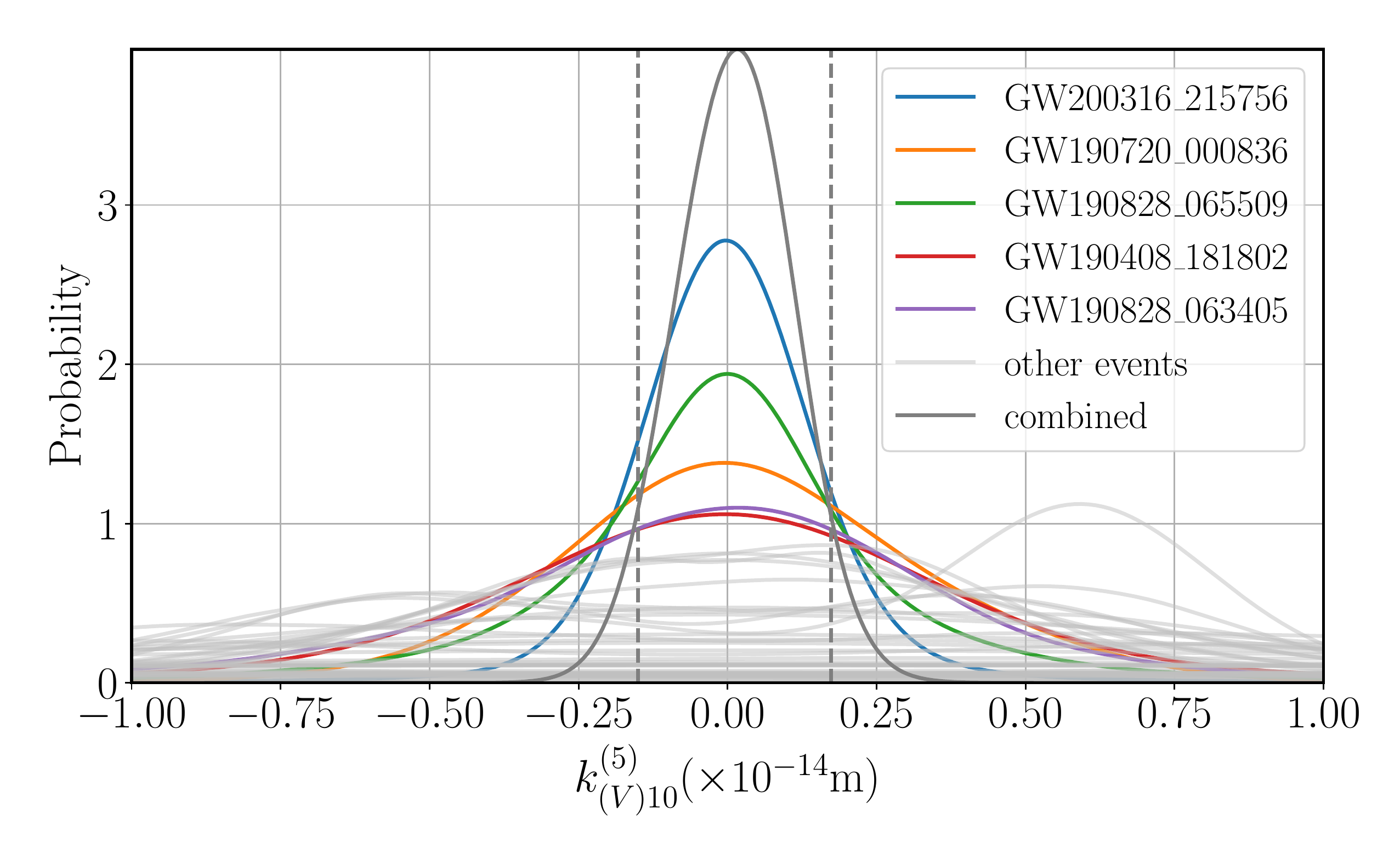}
    \includegraphics[width=0.45\columnwidth]{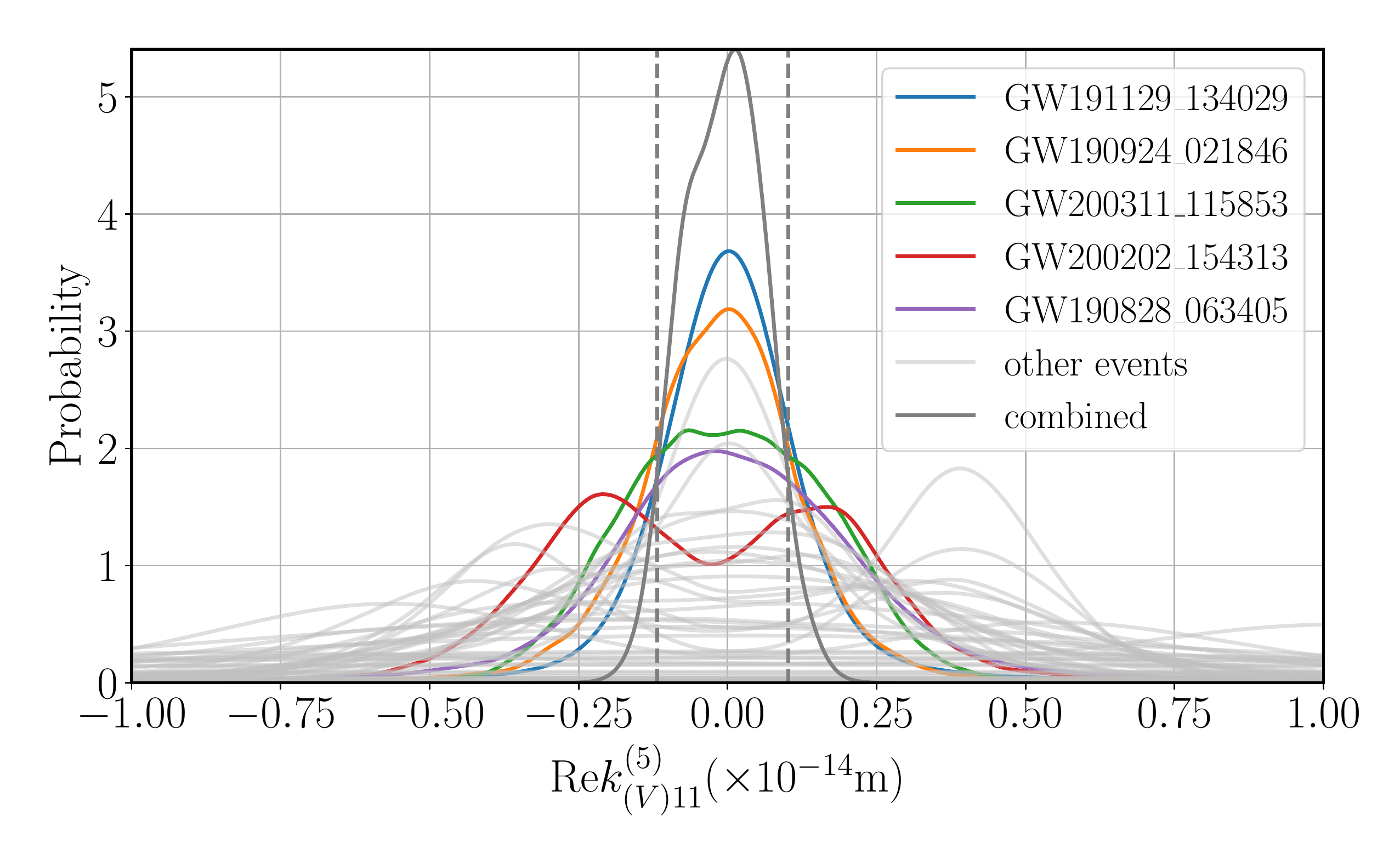}
    \includegraphics[width=0.45\columnwidth]{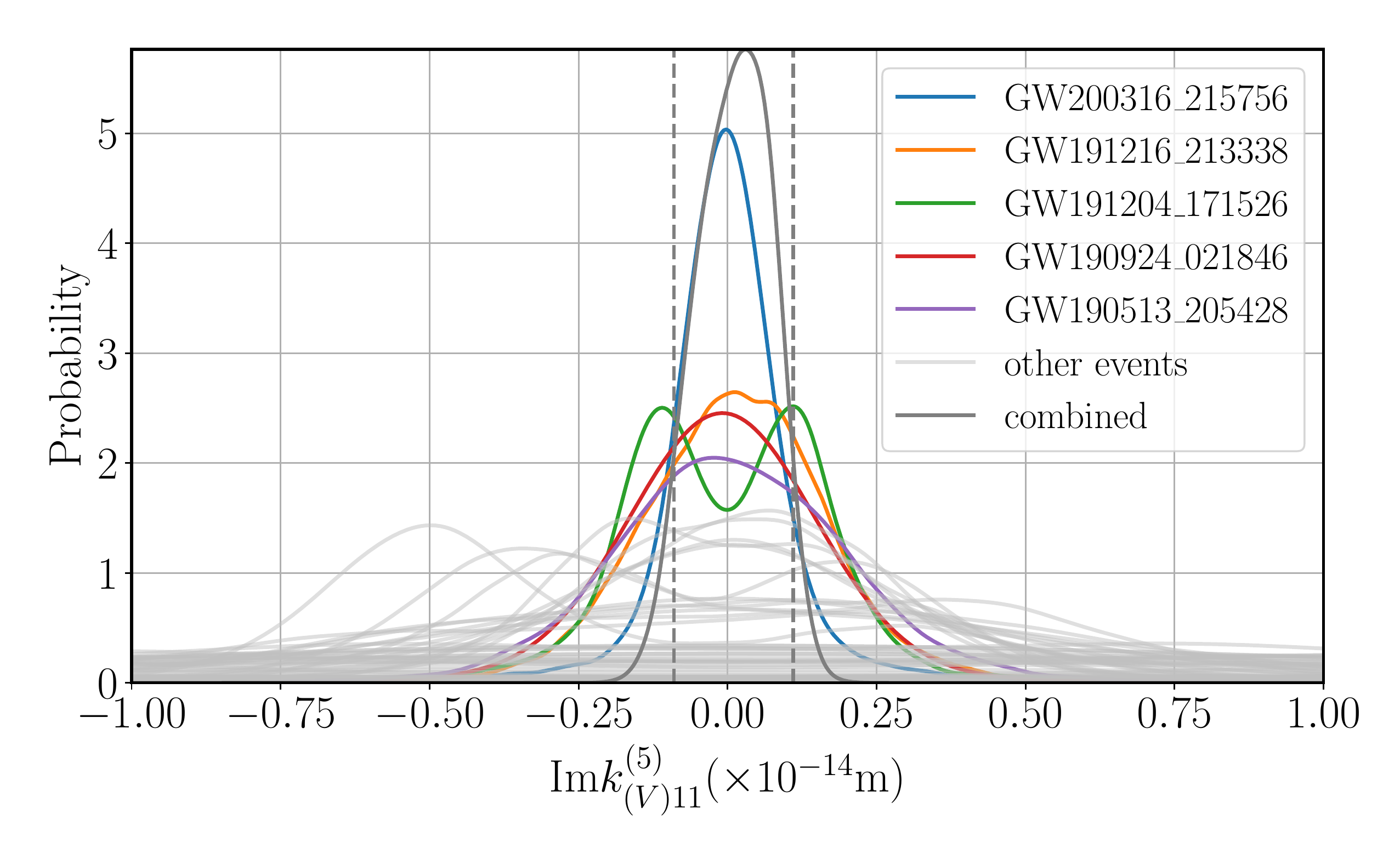}
    \includegraphics[width=0.45\columnwidth]{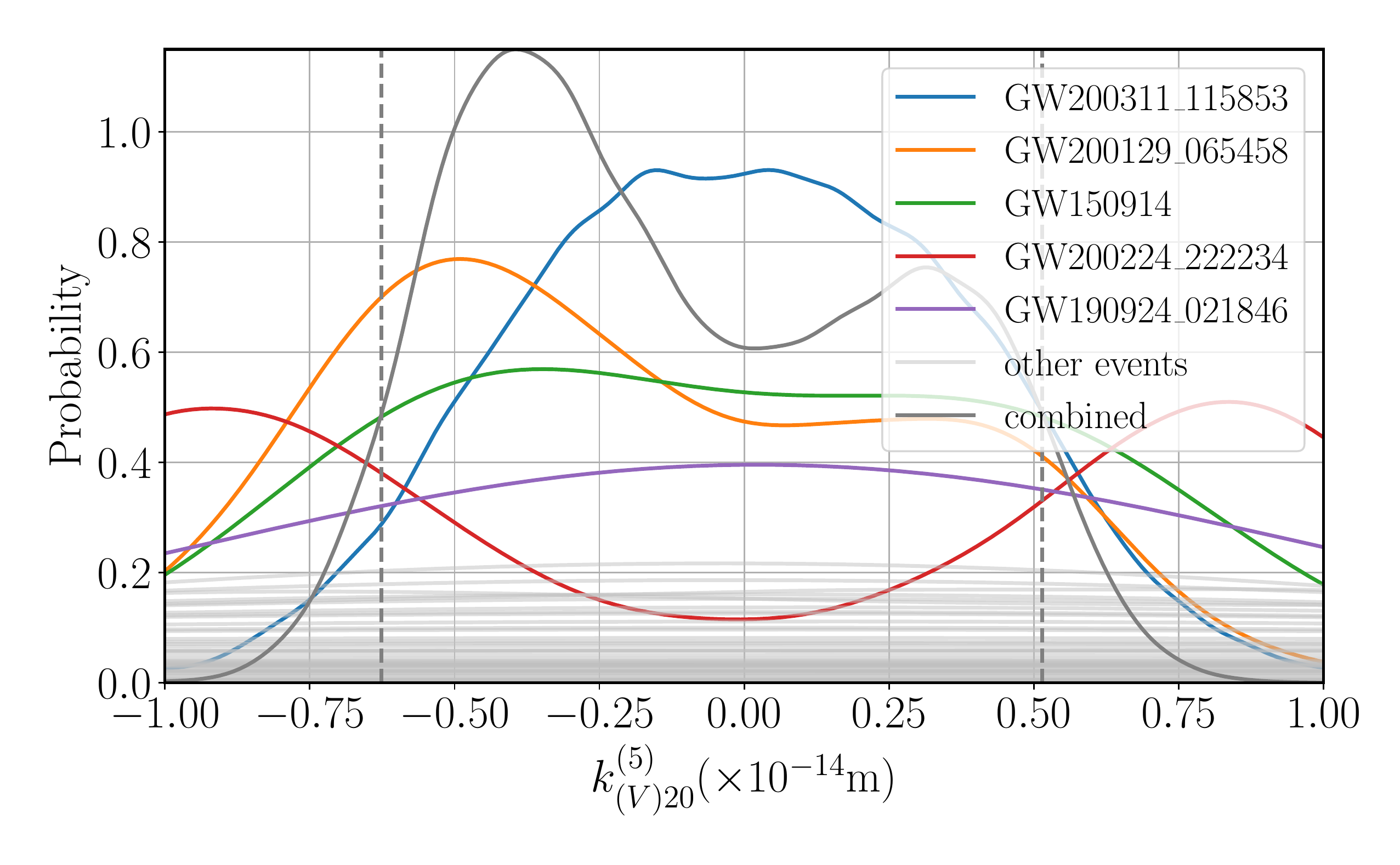}
    \includegraphics[width=0.45\columnwidth]{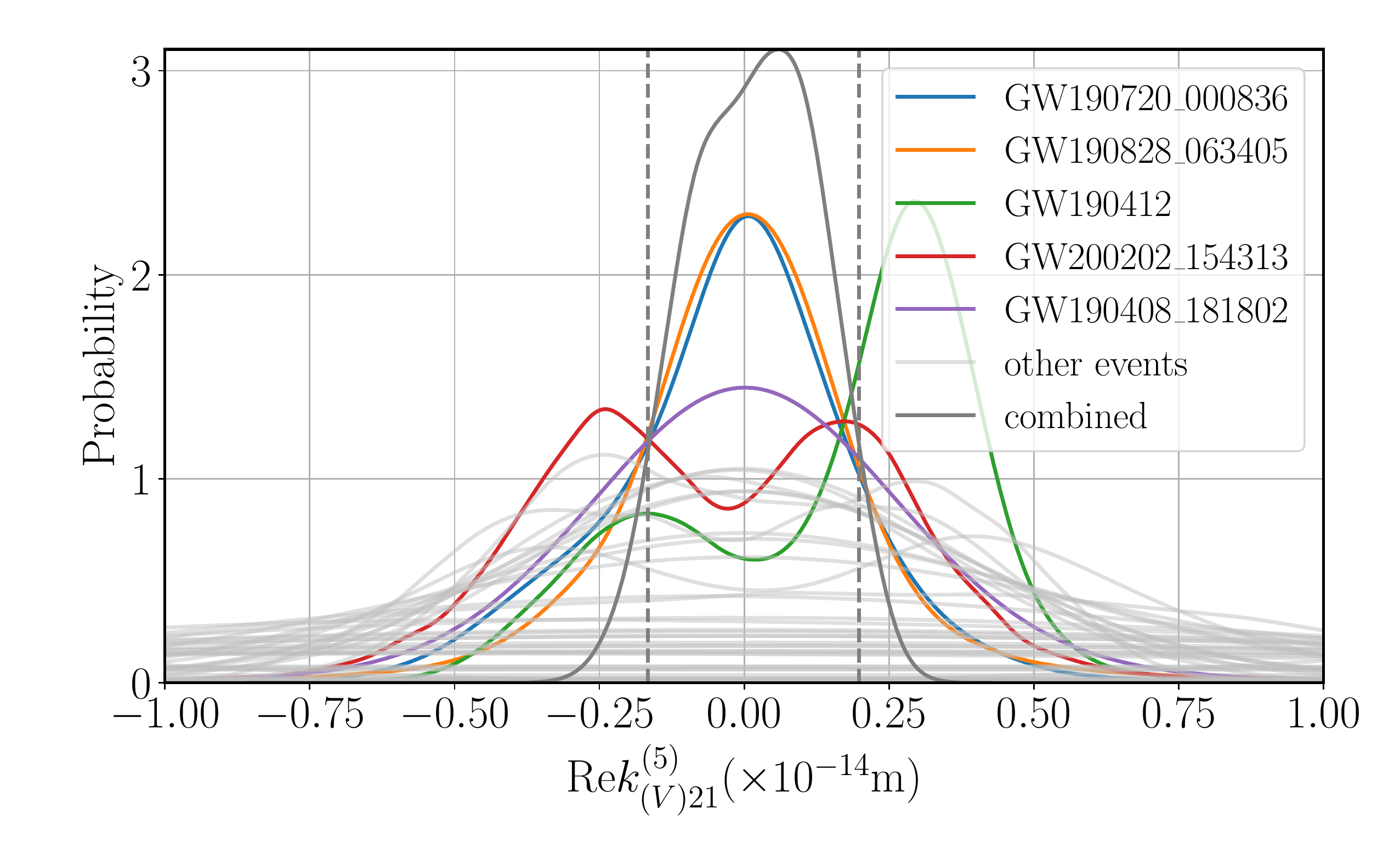}
    \includegraphics[width=0.45\columnwidth]{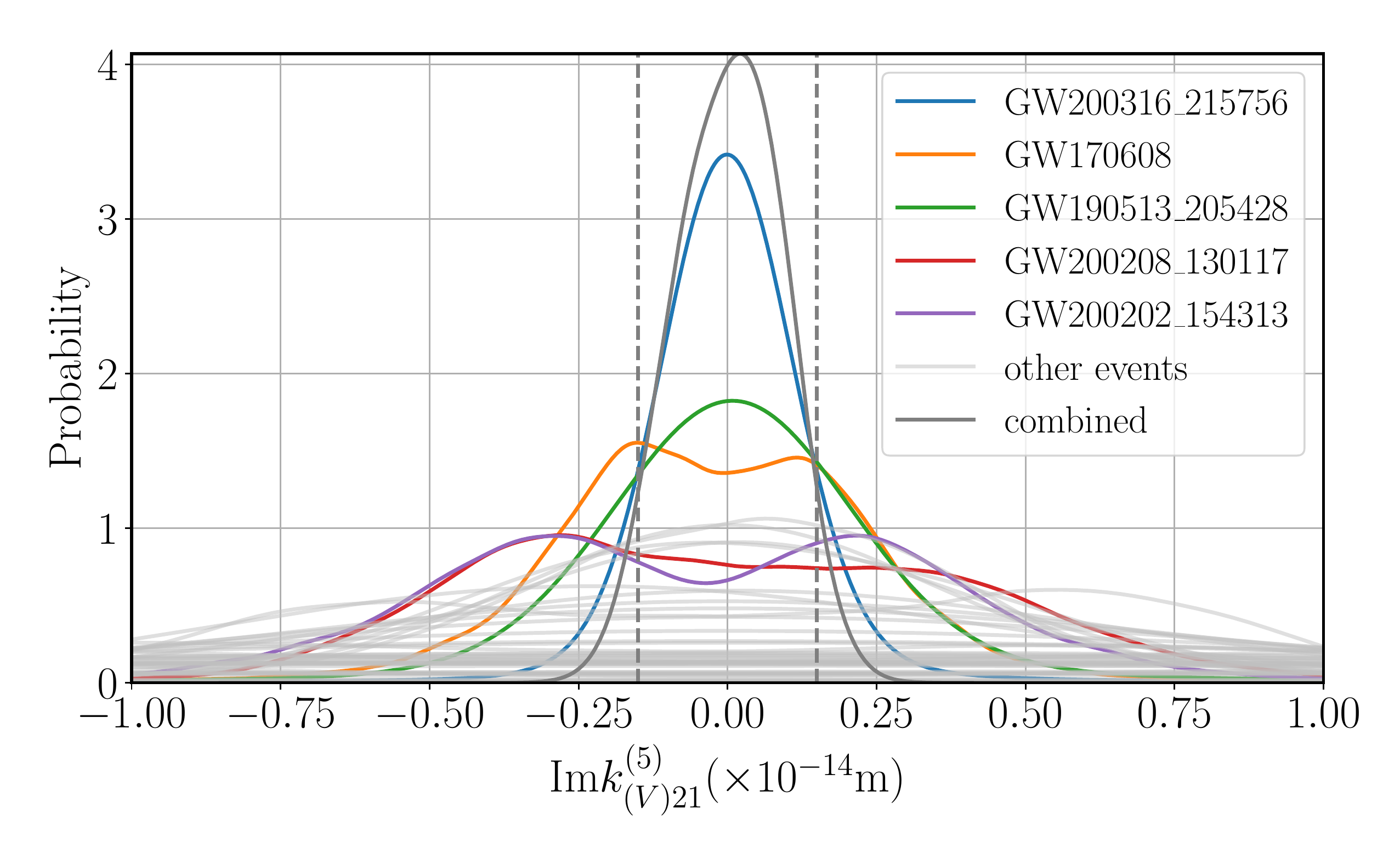}
    \includegraphics[width=0.45\columnwidth]{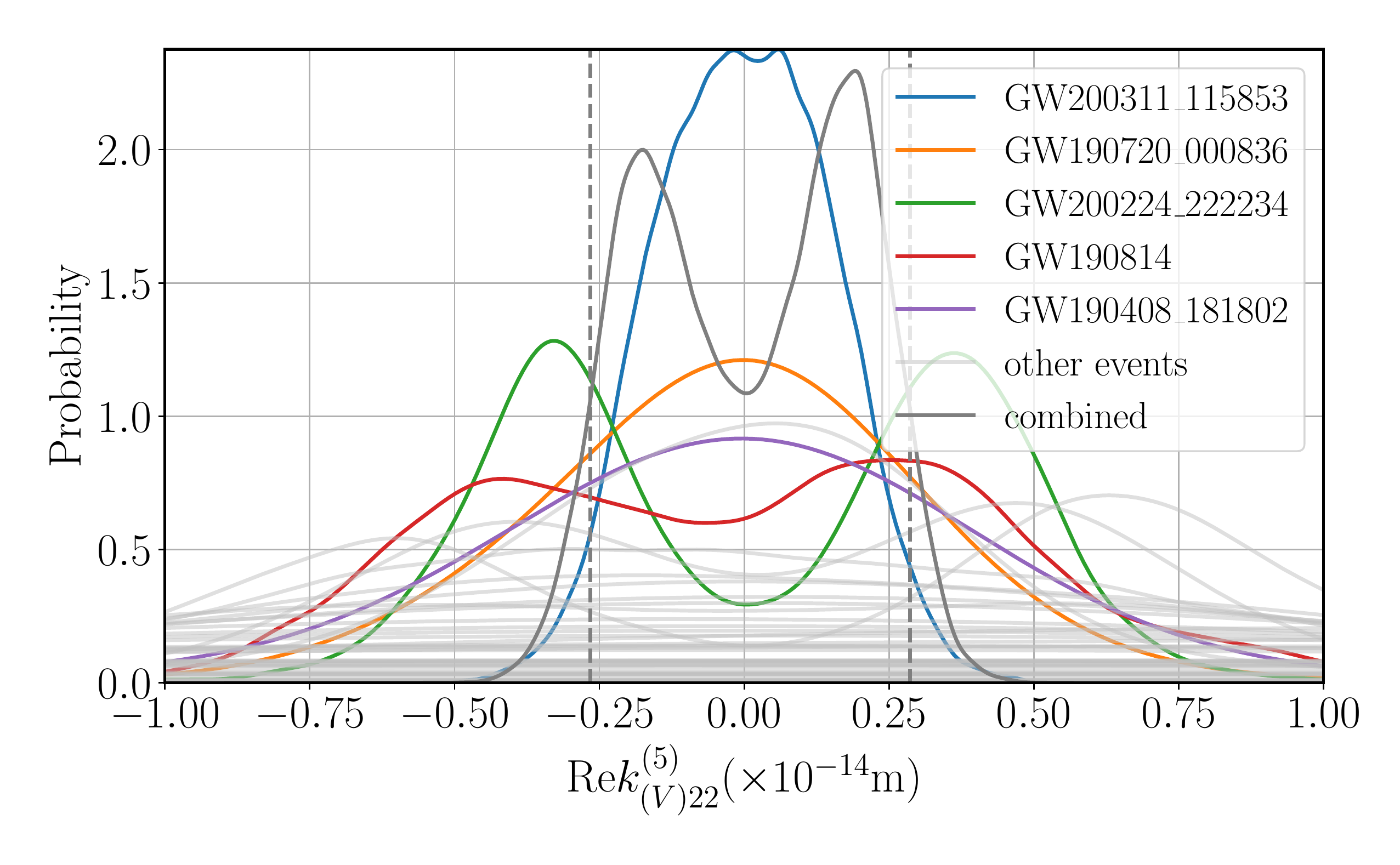}
\end{figure}

\begin{figure}[p]
    \centering
    \includegraphics[width=0.45\columnwidth]{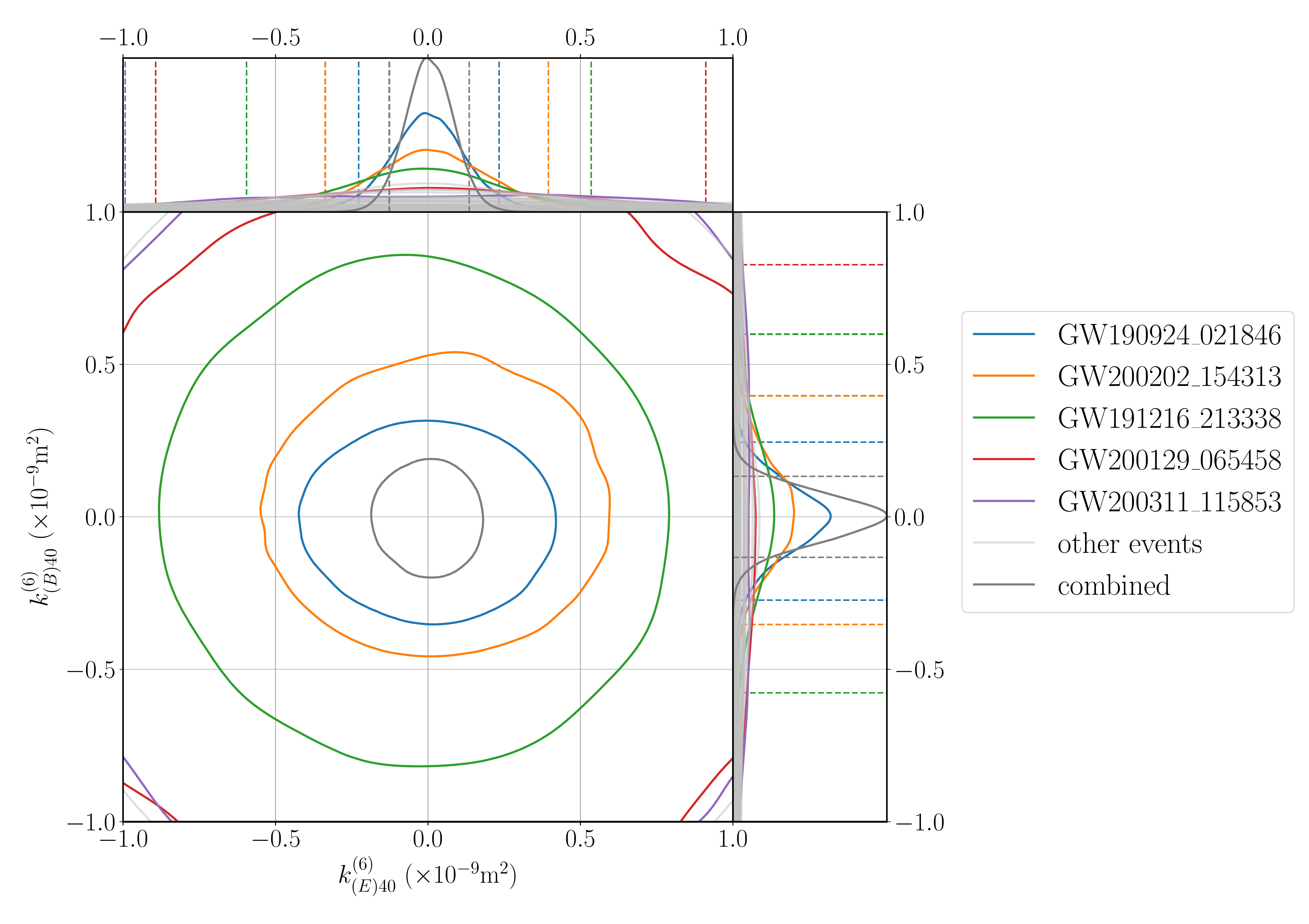}
    \includegraphics[width=0.45\columnwidth]{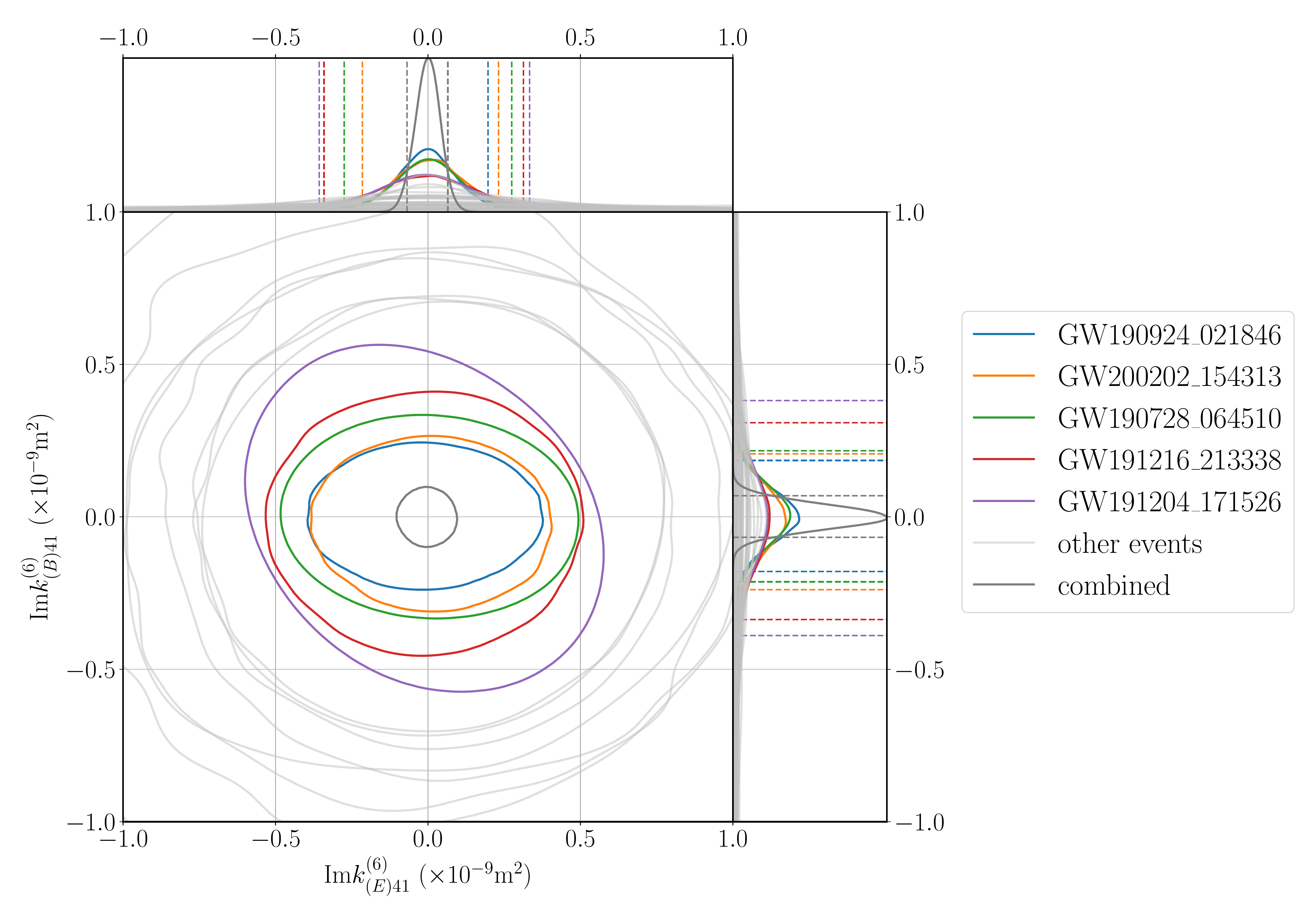}
    \includegraphics[width=0.45\columnwidth]{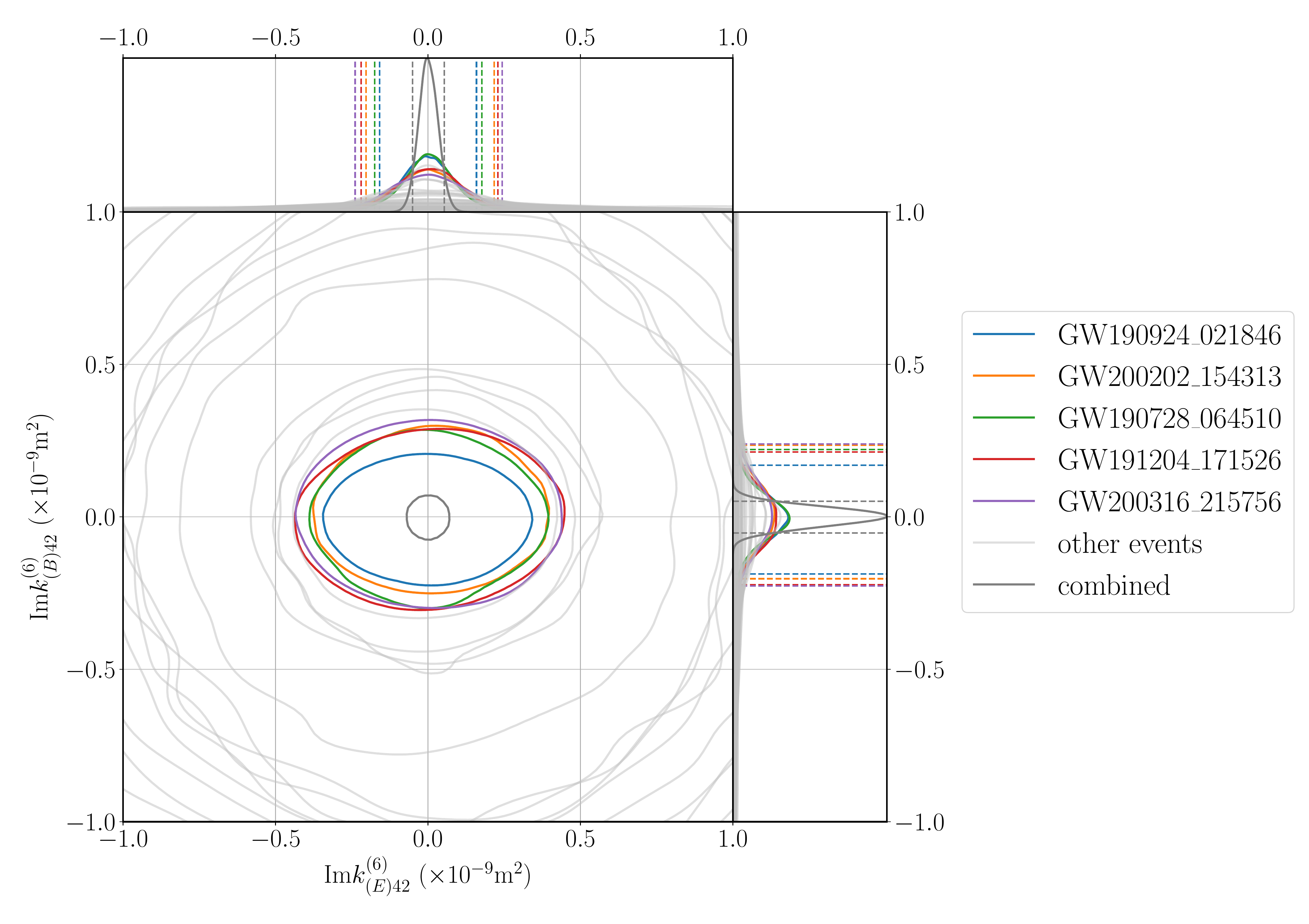}
    \includegraphics[width=0.45\columnwidth]{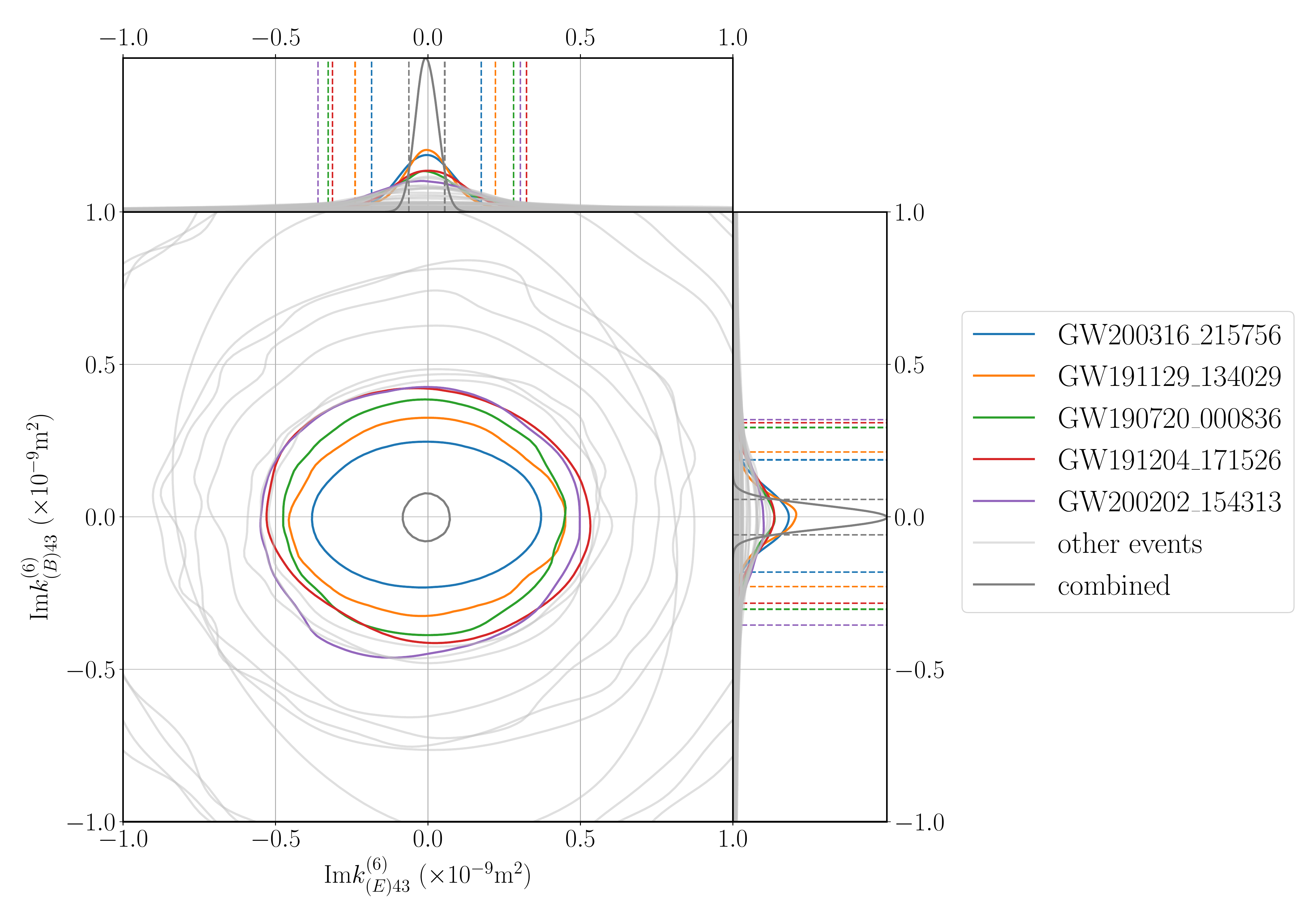}
    \includegraphics[width=0.45\columnwidth]{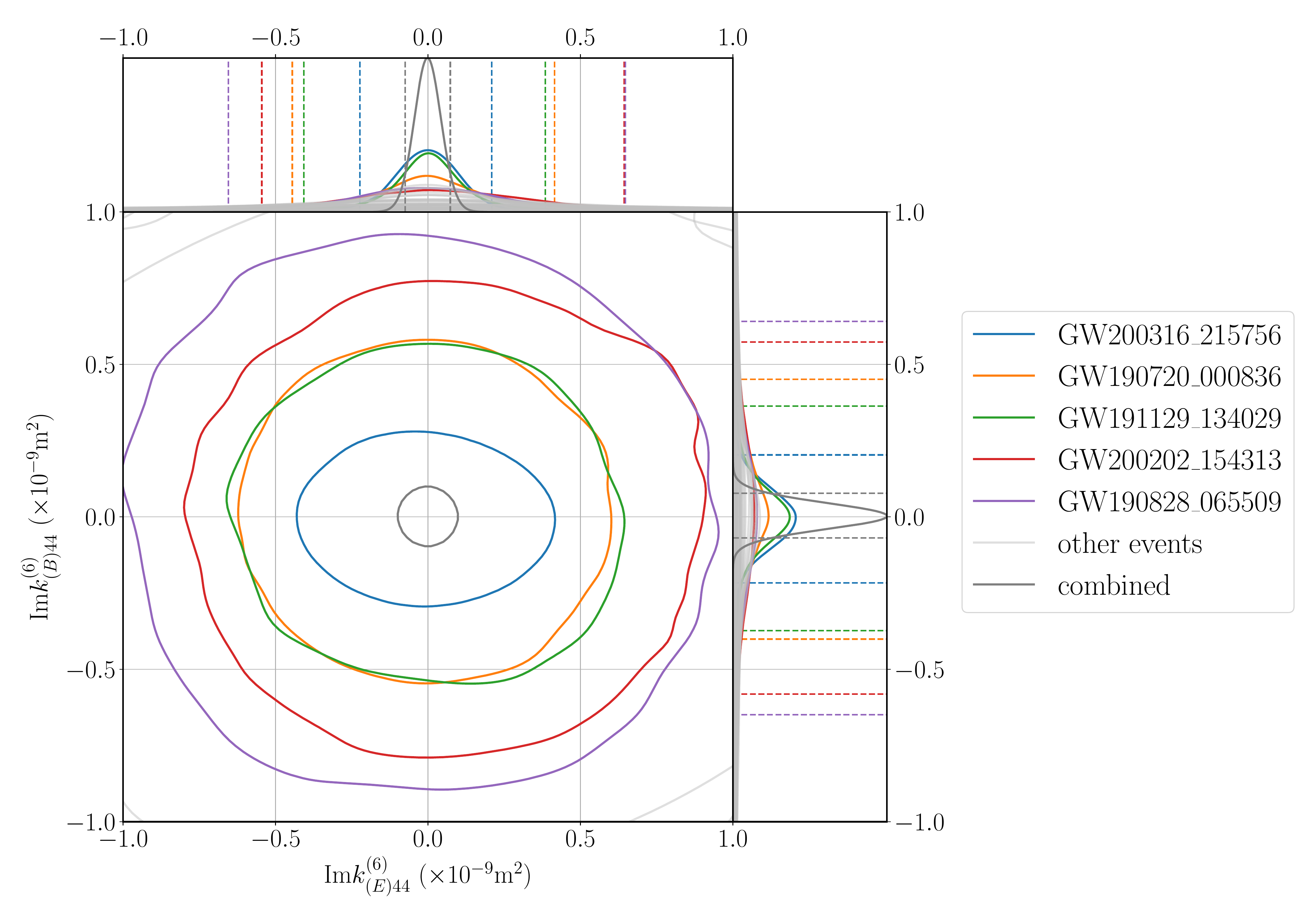}
    \includegraphics[width=0.45\columnwidth]{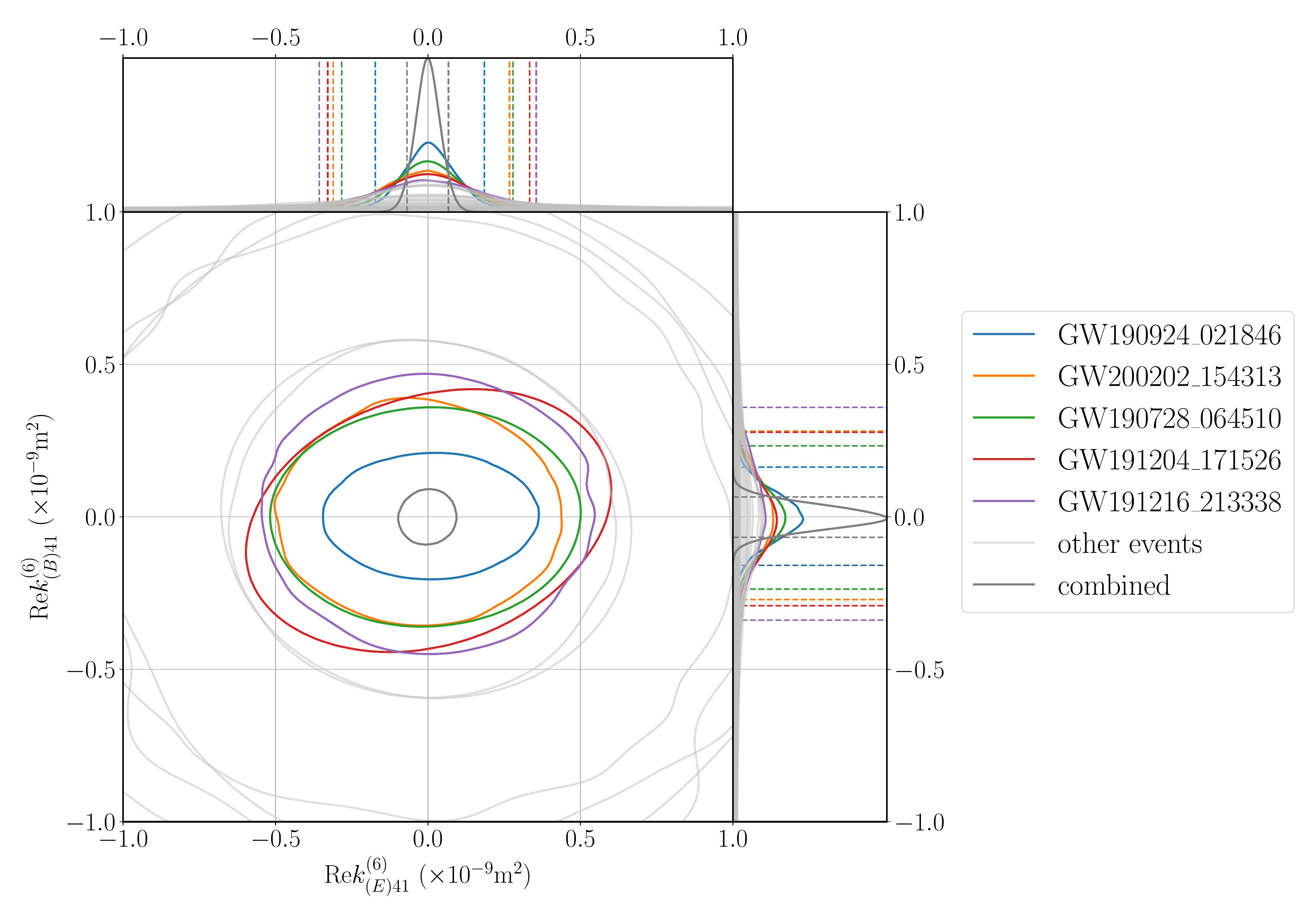}
\end{figure}

\newpage

\begin{table}
\centering

\resizebox{0.95\columnwidth}{!}{
\begin{tabular}{lccccccc}
    \toprule
    Events               &       $\mathcal{M}(M_\odot)$     &        $m_1(M_\odot)$            &     $m_2(M_\odot)$             &    $D_\mathrm{L}(\mathrm{Mpc})$ &      SNR                    &     FAR$(\mathrm{yr}^{-1})$  &     $p_{\mathrm{astro}}$ \\
    \midrule
    GW150914             &       $28.6_{-1.5}^{+1.7}$       &      $35.6_{-3.1}^{+4.7}$        &      $30.6_{-4.4}^{+3.0}$      &    $440_{-170}^{+150}$        &      $24.4$                   &         $10^{-7}$        &      $1.0$ \\
    GW151226             &       $8.9_{-0.3}^{+0.3}$        &      $13.7_{-3.2}^{+8.8}$        &      $7.7_{-2.5}^{+2.2}$       &    $450_{-190}^{+180}$        &      $13.1$                   &         $10^{-7}$        &      $1.0$ \\
    GW170104             &       $21.4_{-1.8}^{+2.2}$       &      $30.8_{-5.6}^{+7.3}$        &      $20.0_{-4.6}^{+4.9}$      &    $990_{-430}^{+440}$        &      $13.0$                   &         $10^{-7}$        &      $1.0$ \\
    GW170608             &       $7.9_{-0.2}^{+0.2}$        &      $11.0_{-1.7}^{+5.5}$        &      $7.6_{-2.2}^{+1.4}$       &    $320_{-110}^{+120}$        &      $14.9$                   &         $10^{-7}$        &      $1.0$ \\
    GW170809             &       $24.9_{-1.7}^{+2.1}$       &      $35.0_{-5.9}^{+8.3}$        &      $23.8_{-5.2}^{+5.1}$      &    $1030_{-390}^{+320}$       &      $12.4$                   &         $10^{-7}$        &      $1.0$ \\
    GW170814             &       $24.1_{-1.1}^{+1.4}$       &      $30.6_{-3.0}^{+5.6}$        &      $25.2_{-4.0}^{+2.8}$      &    $600_{-220}^{+150}$        &      $15.9$                   &         $10^{-7}$        &      $1.0$ \\
    ${\rm GW170817^*}$         &       $1.186_{-0.001}^{+0.001}$  &      $1.46_{-0.1}^{+0.12}$       &      $1.27_{-0.09}^{+0.09}$    &    $40_{-15}^{+7}$            &      $33.0$                   &         $10^{-7}$        &      $1.0$ \\
    GW170818             &       $26.5_{-1.7}^{+2.1}$       &      $35.4_{-4.7}^{+7.5}$        &      $26.7_{-5.2}^{+4.3}$      &    $1060_{-380}^{+420}$       &      $11.3$                   &    $4.2\times10^{-5}$     &      $1.0$ \\
    GW170823             &       $29.2_{-3.6}^{+4.6}$       &      $39.5_{-6.7}^{+11.2}$       &      $29.0_{-7.8}^{+6.7}$      &    $1940_{-900}^{+970}$       &      $11.5$                   &         $10^{-7}$        &      $1.0$ \\
    GW190408\_181802     &       $18.3_{-1.2}^{+1.9}$       &      $24.6_{-3.4}^{+5.1}$        &      $18.4_{-3.6}^{+3.3}$      &    $1550_{-600}^{+400}$       &      $14.6$                   &         $10^{-5}$        &      $1.0$ \\
    GW190412             &       $13.3_{-0.3}^{+0.4}$       &      $30.1_{-5.1}^{+4.7}$        &      $8.3_{-0.9}^{+1.6}$       &    $740_{-170}^{+140}$        &      $18.8$                   &         $10^{-5}$        &      $1.0$ \\
    GW190421\_213856     &       $31.2_{-4.2}^{+5.9}$       &      $41.3_{-6.9}^{+10.4}$       &      $31.9_{-8.8}^{+8.0}$      &    $2880_{-1380}^{+1370}$     &      $10.5$                   &    $7.7\times10^{-4}$     &      $0.99$ \\
    ${\rm GW190425^*}$             &       $1.44_{-0.02}^{+0.02}$     &      $2.0_{-0.3}^{+0.6}$         &      $1.4_{-0.3}^{+0.3}$       &    $160_{-70}^{+70}$          &      $13.0$                   &    $7.5\times10^{-4}$     &        - \\
    ${\rm GW190503\_185404^*}$     &       $30.2_{-4.2}^{+4.2}$       &      $43.3_{-8.1}^{+9.2}$        &      $28.4_{-8.0}^{+7.7}$      &    $1450_{-630}^{+690}$       &      $12.0$                   &         $10^{-5}$        &      $0.99$ \\
    GW190512\_180714     &       $14.6_{-1.0}^{+1.3}$       &      $23.3_{-5.8}^{+5.3}$        &      $12.6_{-2.5}^{+3.6}$      &    $1430_{-550}^{+550}$       &      $12.2$                   &         $10^{-5}$        &      $1.0$ \\
    ${\rm GW190513\_205428^*}$     &       $21.6_{-1.9}^{+3.8}$       &      $35.7_{-9.2}^{+9.5}$        &      $18.0_{-4.1}^{+7.7}$      &    $2060_{-800}^{+880}$       &      $12.2$                   &         $10^{-5}$        &      $1.0$ \\
    GW190517\_055101     &       $26.6_{-4.0}^{+4.0}$       &      $37.4_{-7.6}^{+11.7}$       &      $25.3_{-7.3}^{+7.0}$      &    $1860_{-840}^{+1620}$      &      $10.2$                   &    $5.7\times10^{-5}$     &      $1.0$ \\
    GW190519\_153544     &       $44.5_{-7.1}^{+6.4}$       &      $66.0_{-12.0}^{+10.7}$      &      $40.5_{-11.1}^{+11.0}$    &    $2530_{-920}^{+1830}$      &      $12.0$                   &         $10^{-5}$        &      $0.99$ \\
    GW190521             &       $69.2_{-10.6}^{+17.0}$     &      $95.3_{-18.9}^{+28.7}$      &      $69.0_{-23.1}^{+22.7}$    &    $3920_{-1950}^{+2190}$     &      $14.3$                   &    $2.0\times10^{-4}$     &        - \\
    GW190521\_074359     &       $32.1_{-2.5}^{+3.2}$       &      $42.2_{-4.8}^{+5.9}$        &      $32.8_{-6.4}^{+5.4}$      &    $1240_{-570}^{+400}$       &      $24.3$                   &         $10^{-5}$        &      $1.0$ \\
    GW190602\_175927     &       $49.1_{-8.5}^{+9.1}$       &      $69.1_{-13.0}^{+15.7}$      &      $47.8_{-17.4}^{+14.3}$    &    $2690_{-1120}^{+1790}$     &      $12.1$                   &    $1.1\times10^{-5}$     &      $0.99$ \\
    GW190630\_185205     &       $24.9_{-2.1}^{+2.1}$       &      $35.1_{-5.6}^{+6.9}$        &      $23.7_{-5.1}^{+5.2}$      &    $890_{-370}^{+560}$        &      $15.6$                   &         $10^{-5}$        &      $1.0$ \\
    GW190706\_222641     &       $42.7_{-7.0}^{+10.0}$      &      $67.0_{-16.2}^{+14.6}$      &      $38.2_{-13.3}^{+14.6}$    &    $4420_{-1930}^{+2590}$     &      $12.3$                   &         $10^{-5}$        &      $1.0$ \\
    GW190707\_093326     &       $8.5_{-0.5}^{+0.6}$        &      $11.6_{-1.7}^{+3.3}$        &      $8.4_{-1.7}^{+1.4}$       &    $770_{-370}^{+380}$        &      $12.9$                   &         $10^{-5}$        &      $1.0$ \\
    GW190708\_232457     &       $13.2_{-0.6}^{+0.9}$       &      $17.6_{-2.3}^{+4.7}$        &      $13.2_{-2.7}^{+2.0}$      &    $880_{-390}^{+330}$        &      $13.0$                   &    $2.8\times10^{-5}$     &      $0.99$ \\
    GW190720\_000836     &       $8.9_{-0.8}^{+0.5}$        &      $13.4_{-3.0}^{+6.7}$        &      $7.8_{-2.2}^{+2.3}$       &    $790_{-320}^{+690}$        &      $11.6$                   &         $10^{-5}$        &      $1.0$ \\
    GW190727\_060333     &       $28.6_{-3.7}^{+5.3}$       &      $38.0_{-6.2}^{+9.5}$        &      $29.4_{-8.4}^{+7.1}$      &    $3300_{-1500}^{+1540}$     &      $12.2$                   &         $10^{-5}$        &      $1.0$ \\
    GW190728\_064510     &       $8.6_{-0.3}^{+0.5}$        &      $12.3_{-2.2}^{+7.2}$        &      $8.1_{-2.6}^{+1.7}$       &    $870_{-370}^{+260}$        &      $13.6$                   &         $10^{-5}$        &      $1.0$ \\
    GW190814             &       $6.09_{-0.06}^{+0.06}$     &      $23.2_{-1.0}^{+1.1}$        &      $2.59_{-0.09}^{+0.08}$    &    $240_{-50}^{+40}$          &      $22.1$                   &         $10^{-5}$        &      $1.0$ \\
    GW190828\_063405     &       $25.0_{-2.1}^{+3.4}$       &      $32.1_{-4.0}^{+5.8}$        &      $26.2_{-4.8}^{+4.6}$      &    $2130_{-930}^{+660}$       &      $16.0$                   &         $10^{-5}$        &      $1.0$ \\
    GW190828\_065509     &       $13.3_{-1.0}^{+1.2}$       &      $24.1_{-7.2}^{+7.0}$        &      $10.2_{-2.1}^{+3.6}$      &    $1600_{-600}^{+620}$       &      $11.1$                   &         $10^{-5}$        &      $1.0$ \\
    GW190910\_112807     &       $34.3_{-4.1}^{+4.1}$       &      $43.9_{-6.1}^{+7.6}$        &      $35.6_{-7.2}^{+6.3}$      &    $1460_{-580}^{+1030}$      &      $13.4$                   &    $1.9\times10^{-5}$     &      $0.99$ \\
    GW190915\_235702     &       $25.3_{-2.7}^{+3.2}$       &      $35.3_{-6.4}^{+9.5}$        &      $24.4_{-6.1}^{+5.6}$      &    $1620_{-610}^{+710}$       &      $13.0$                   &         $10^{-5}$        &      $1.0$ \\
    ${\rm GW190924\_021846^*}$     &       $5.8_{-0.2}^{+0.2}$        &      $8.9_{-2.0}^{+7.0}$         &      $5.0_{-1.9}^{+1.4}$       &    $570_{-220}^{+220}$        &      $13.1$     &         $10^{-5}$        &      $1.0$ \\
    ${\rm GW191109\_010717^*}$     &       $47.5_{-7.5}^{+9.6}$       &      $65.0_{-11.0}^{+11.0}$      &      $47.0_{-13.0}^{+15.0}$    &    $1290_{-650}^{+1130}$      &      $17.3$     &    $1.8\times10^{-4}$     &      $0.99$ \\
    GW191129\_134029     &       $7.31_{-0.28}^{+0.43}$     &      $10.7_{-2.1}^{+4.1}$        &      $6.7_{-1.7}^{+1.5}$       &    $790_{-330}^{+260}$        &      $13.1$     &         $10^{-5}$        &      $0.99$ \\
    GW191204\_171526     &       $8.55_{-0.27}^{+0.38}$     &      $11.9_{-1.8}^{+3.3}$        &      $8.2_{-1.6}^{+1.4}$       &    $650_{-250}^{+190}$        &      $17.5$     &         $10^{-5}$        &      $0.99$ \\
    GW191215\_223052     &       $18.4_{-1.7}^{+2.2}$       &      $24.9_{-4.1}^{+7.1}$        &      $18.1_{-4.1}^{+3.8}$      &    $1930_{-860}^{+890}$       &      $11.2$     &         $10^{-5}$        &      $0.99$ \\
    GW191216\_213338     &       $8.33_{-0.19}^{+0.22}$     &      $12.1_{-2.3}^{+4.6}$        &      $7.7_{-1.9}^{+1.6}$       &    $340_{-130}^{+120}$        &      $18.6$     &         $10^{-5}$        &      $0.99$ \\
    GW191222\_033537     &       $33.8_{-5.0}^{+7.1}$       &      $45.1_{-8.0}^{+10.9}$       &      $34.7_{-10.5}^{+9.3}$     &    $3000_{-1700}^{+1700}$     &      $12.5$     &         $10^{-5}$        &      $0.99$ \\
    GW200112\_155838     &       $27.4_{-2.1}^{+2.6}$       &      $35.6_{-4.5}^{+6.7}$        &      $28.3_{-5.9}^{+4.4}$      &    $1250_{-460}^{+430}$       &      $19.8$     &         $10^{-5}$        &      $0.99$ \\
    ${\rm GW200115\_042309^*}$     &       $2.43_{-0.07}^{+0.05}$     &      $5.9_{-2.5}^{+2.0}$         &      $1.44_{-0.29}^{+0.85}$    &    $290_{-100}^{+150}$        &      $11.3$     &         $10^{-5}$        &      $0.99$ \\
    ${\rm GW200129\_065458^*}$     &       $27.2_{-2.3}^{+2.1}$       &      $34.5_{-3.2}^{+9.9}$        &      $28.9_{-9.3}^{+3.4}$      &    $900_{-380}^{+290}$        &      $26.8$     &         $10^{-5}$        &      $0.99$ \\
    GW200202\_154313     &       $7.49_{-0.2}^{+0.24}$      &      $10.1_{-1.4}^{+3.5}$        &      $7.3_{-1.7}^{+1.1}$       &    $410_{-160}^{+150}$        &      $10.8$     &         $10^{-5}$        &      $0.99$ \\
    GW200208\_130117     &       $27.7_{-3.1}^{+3.6}$       &      $37.8_{-6.2}^{+9.2}$        &      $27.4_{-7.4}^{+6.1}$      &    $2230_{-850}^{+1000}$      &      $10.8$     &     $3.1\times10^{-4}$    &      $0.99$ \\
    GW200219\_094415     &       $27.6_{-3.8}^{+5.6}$       &      $37.5_{-6.9}^{+10.1}$       &      $27.9_{-8.4}^{+7.4}$      &    $3400_{-1500}^{+1700}$     &      $10.7$     &     $9.9\times10^{-4}$    &      $0.99$ \\
    GW200224\_222234     &       $31.1_{-2.6}^{+3.2}$       &      $40.0_{-4.5}^{+6.9}$        &      $32.5_{-7.2}^{+5.0}$      &    $1710_{-640}^{+490}$       &      $20.0$     &         $10^{-5}$        &      $0.99$ \\
    GW200225\_060421     &       $14.2_{-1.4}^{+1.5}$       &      $19.3_{-3.0}^{+5.0}$        &      $14.0_{-3.5}^{+2.8}$      &    $1150_{-530}^{+510}$       &      $12.5$     &         $1.1\times10^{-5}$       &      $0.99$ \\
    GW200311\_115853     &       $26.6_{-2.0}^{+2.4}$       &      $34.2_{-3.8}^{+6.4}$        &      $27.7_{-5.9}^{+4.1}$      &    $1170_{-400}^{+280}$       &      $17.8$     &         $10^{-5}$        &      $0.99$ \\
    GW200316\_215756     &       $8.75_{-0.55}^{+0.62}$     &      $13.1_{-2.9}^{+10.2}$       &      $7.8_{-2.9}^{+1.9}$       &    $1120_{-440}^{+470}$       &      $10.3$     &         $10^{-5}$        &      $0.99$ \\
    \bottomrule
\end{tabular}
}
\caption{{\bf Events used in the analysis.} Data are from \href{https://www.gw-openscience.org}{Gravitational Wave Open Science Center}. Values in the table are from default search and default parameter estimation. Error bars on SNR are not shown here for neatness and clarity, and $10^{-5}$ or $10^{-7}$ are upper bounds of FAR. See instructions in above website for more details. Asterisked events denote that glitch subtracted data are used in parameter estimation.}
\label{tab_all_events}
\end{table}

\end{document}